# A TRANSFORMATION–BASED APPROACH FOR THE DESIGN OF PARALLEL/DISTRIBUTED SCIENTIFIC SOFTWARE: THE FFT [*]

HARRY B. HUNT , LENORE R. MULLIN , DANIEL J. ROSENKRANTZ [†], AND JAMES E. RAYNOLDS[‡]

**Abstract.** We describe a methodology for designing efficient parallel and distributed scientific software. This methodology utilizes sequences of mechanizable algebra–based optimizing transformations. In this study, we apply our methodology to the FFT, starting from a high–level algebraic algorithm description. Abstract multiprocessor plans are developed and refined to specify which computations are to be done by each processor. Templates are then created that specify the locations of computations and data on the processors, as well as data flow among processors. Templates are developed in both the MPI and OpenMP programming styles.

Preliminary experiments comparing code constructed using our methodology with code from several standard scientific libraries show that our code is often competitive and sometimes performs better. Interestingly, our code handled a larger range of problem sizes on one target architecture.

## 1. Introduction. —

We present a systematic, algebraically based, design methodology for efficient implementation of computer programs optimized over multiple levels of the processor/memory and network hierarchy. Using a common formalism to describe the problem and the partitioning of data over processors and memory levels allows one to mathematically prove the efficiency and correctness of a given algorithm as measured in terms of a set of metrics (such as processor/network speeds, etc.). The approach allows the average programmer to achieve high-level optimizations similar to those used by compiler writers (e.g. the notion of *tiling*).

The approach is similar in spirit to other efforts using libraries of algorithm building blocks based on C++ template classes. In POOMA for example, expression templates using the Portable Expression Template Engine (PETE) (http://www.acl.lanl.gov.pete) were used to achieve efficient distribution of array indexing over scalar operations [1, 2, 3, 4, 5, 6, 7, 8, 9, 10].

As another example, The Matrix Template Library (MTL) [11, 12] is a system that handles dense and sparse matrices, and uses template meta-programs to generate tiled algorithms for dense matrices.

For example:

$$(A + B)_{ij} = A_{ij} + B_{ij}, \qquad (1.1)$$

can be generalized to the situation in which the multi-dimensional arrays $A$ and $B$ are selected using a vector of indices $\vec{v}$.

In POOMA $A$ and $B$ were represented as classes and expression templates were used as re-write rules to efficiently carry out the translation to scalar operations implied by:

[*]RESEARCH SUPPORTED BY NSF GRANT CCR–0105536.

[†]Department of Computer Science, University at Albany, State University of New York, Albany, NY 12203

[‡]College of Nanoscale Science and Engineering, University at Albany, State University of New York, Albany, NY 12203

$$(A + B)_{\vec{v}} = A_{\vec{v}} + B_{\vec{v}} \tag{1.2}$$

The approach presented in this paper makes use of A Mathematics of Arrays (MoA) and an indexing calculus (i.e. the $\psi$ calculus) to enable the programmer to develop algorithms using high-level compiler-like optimizations through the ability to algebraically compose and reduce sequences of array operations.

As such, the translation from the left hand side of Eq. 1.2 to the right side is just one of a wide variety of operations that can be carried out using this algebra. In the MoA formalism, the array expression in Eq. 1.2 would be written:

$$\vec{v}\psi(A + B) = \vec{v}\psi A + \vec{v}\psi B \tag{1.3}$$

where we have introduced the *psi-operator* $\psi$ to denote the operation of extracting an element from the multi-dimensional array using the index vector ($\vec{v}$).

In this paper we give a demonstration of the approach as applied to the creation of efficient implementations of the Fast Fourier Transform (FFT) optimized over multi-processor, multi-memory/network, environments. Multi-dimensional data arrays are *reshaped* through the process of *dimension lifting* to explicitly add dimensions to enable indexing blocks of data over the various levels of the hierarchy. A sequence of array operations (represented by the various operators of the algebra acting on arrays) is algebraically composed to achieve the *Denotational Normal Form* (DNF). Then the $\psi$-calculus is used to reduce the DNF to the ONF (*Operational Normal Form*) which explicitly expresses the algorithm in terms of loops and operations on indices. The ONF can thus be directly translated into efficient code in the language of the programmer's choice be it for hardware or software application.

The application we use as a demonstration vehicle – the Fast Fourier Transform – is of significant interest in its own right. The Fast Fourier Transform (FFT) is one of the most important computational algorithms and its use is pervasive in science and engineering. The work in this paper is a companion to two previous papers [13] in which the FFT was optimized in terms of in-cache operations leading to factors of *two* to *four* speedup in comparison with our previous records. Further background material including comparisons with library routines can be found in Refs. [14, 15, 16, 17] and [18].

Our algorithm can be seen to be a generalization of similar work aimed at out-of-core optimizations [19]. Similarly, block decompositions of matrices (in general) are special cases of our *rehape-traspose* design. Most importantly, our designs are general for any partition size, i.e. not necessary blocked in squares, and any number of dimensions. Furthermore, our designs use linear transformations from an algebraic specification and thus they are **verified**. Thus, by specifying designs (such as Cormen's and others [19]) using these techniques, these designs too could be verified.

The purpose of this paper **IS NOT** to attempt any serious analysis of the number of cache misses incurred by the algorithm in the spirit of of Hong and Kung and others [20, 21, 22]. Rather, we present an **algebraic** method that achieves (or is competitive) with such optimizations **mechanically**. Through linear transformations we produce a **normal form**, the ONF, that is directly implementable in any hardware or software language and is realized in any of the processor/memory levels [23]. Most importantly, our designs are completely general in that through **dimension lifting** we can produce any number of levels in the processor/memory hierarchy.

One objection to our approach is that one might incur an unacceptable performance cost due to the periodic rearrangement of the data that is often needed. This will not, however, be the case if we pre-fetch data before it is needed. The necessity to pre-fetch data also exists in other similar cache-optimized schemes. Our algorithm does what the compiler community calls *tiling*. Since we have analyzed the loop structures, access patterns, and speeds of the processor/memory levels, pre-fetching becomes a deterministic cost function that can easily be combined with *reshape-transpose* or *tiling* operations.

Again we make no attempt to optimize the algorithm for any particular architecture. We provide a general algorithm in the form of an Operational Normal Form that allows the user to specify the blocking size at run time. This ONF therefore enables the individual user to choose the blocking size that gives the best performance for any individual machine, assuming this intentional information can be processed by a compiler[1].

It is also important to note the importance of running *reproducible and deterministic* experiments. Such experiments are only possible when dedicated resources exist *AND* no interrupts or randomness effects memory/cache/communications behavior. This means that multiprocessing and time sharing must be turned off for both OS's and Networks.

Conformal Computing[2] is the name given by the authors to this algebraic approach to the construction of computer programs for array-based computations in science and engineering. The reader should not be misled by the name. Conformal in this sense is not related to *Conformal Mapping* or similar constructs from mathematics although it was inspired by these concepts in which certain properties are preserved under transformations. In particular, by *Conformal Computing* we mean a mathematical system that *conforms* as closely as possible to the underlying structure of the hardware. Further details of the theory including discussion of MoA and $\psi$-calculus are provided in the appendix.

In this feasibility study, we proceed in a semi–mechanical fashion. At each step of algorithm development, the current version of the algorithm is represented in a code–like form, and a simple basic transformation is applied to this code–like form. Each transformation is easily seen to be mechanizable. We used the algebraic formalism described in Section 1.2 to verify that each transformation produces a semantically equivalent algorithm. At each step, we use judgment in deciding which transformation to carry out. This judgment is based on an understanding of the goals of the transformations.

The following is a more detailed description of our methodology, as carried–out in this feasibility study:

**1.1. Overview of Methodology. Phase 1.** Obtain a high–level description of the problem. In this study, a MATLAB–like description is used.

**Phase 2.** Apply a sequence of transformations to optimize the sequential version of the algorithm. The transformations used are of an algebraic nature, can be interpreted in terms of operations on whole arrays, and should not negatively effect subsequent parallelization.

**Phase 3.** Apply a sequence of transformations to develop a *parallel computation plan*. Such plans consists of sequential code that indicates which parts of the overall work are to be done by each individual processor. For each iteration of an outer loop, the plan specifies which iterations of inner loops should be performed by each of the processors in a multiprocessor system.

---

[1] Processing intentional information will be the topic of a future paper
[2] The name Conformal Computing © is protected. Copyright 2003, The Research Foundation of State University of New York, University at Albany.

**Phase 4.** Given a parallel computation plan, apply a sequence of transformations to produce a *parallel computation template*. Such templates specify a parallel or distributed version of the algorithm by indicating (1) which parts of the overall work are to be done by each individual processor, (2) where data is located, and (3) data movement among processors. Various parallel programming language styles can be used to express such templates. In this study, we use both a message passing *per–processor* style, motivated by MPI [24], and an *all–processor* style, motivated by OpenMP [25].

There is also an implicit fifth phase, in which a parallel computation template is transformed into code expressed in a given programming language.

In the future, we envision that scientific programs will be developed in interactive development environments. Such environments will combine human judgment with compiler–like analysis, transformation, and optimization techniques. A programmer will use knowledge of problem semantics to guide the selection of the transformation at each step, but the application of the transformation and verification of its safety will be done mechanically. A given sequence of transformations will stop at a point where a version of the code has been obtained such that subsequent optimizations can be left to an available optimizing compiler. This feasibility study is a first step towards the development of such an interactive program development environment.

**1.2. Program Development Via Use of Array and Indexing Algebra.** Although not discussed further in this paper, Mullin's *Psi Calculus* [26, 27, 28, 29] plays an underlying role in this feasibility study. This calculus provides a unified mathematical framework for representing and studying arrays, array operations, and array indexing. It is especially useful for developing algorithms centered around transformations involving array addressing, decomposition, reshaping, distribution, etc. Each of the algorithm transformations carried out here can be expressed in terms of the Psi Calculus, and we used the Psi Calculus to verify the correctness of program transformations in the development of our FFT algorithms.

**1.3. Related Results.** A general algebraic framework for Fourier and related transforms, including their discrete versions, is discussed in [30]. As discussed in [31, 32] and using this framework, many algorithms for the FFT can be viewed in terms of computing tensor product decompositions of the matrix $B_L$, discussed in Section 2.1 below. Subsequently, a number of additional algorithms for the FFT and related problems have been developed centered around the use of tensor product decompositions [33, 34, 35, 36, 37, 38]. The work done under the acronym FFTW is based on a compiler that generates efficient sequential FFT code that is adapted to a target architecture and specified problem size [39, 40, 41, 42, 43]. A variety of techniques have been used to construct efficient parallel algorithms for the FFT [44, 45, 46, 47, 48, 49].

Previously, we manually developed several distributed algorithms for the FFT [50], experimenting with variants of *bit–reversal*, *weight* computations, *butterfly* computations, and *message generation* mechanisms. In [50], we report on the evaluation of twelve resulting variant programs for the one–dimensional FFT, evaluated by running a consistent set of experiments. In [14, 15], we compare the performance of our programs with that of several other FFT implementations (FFTW, NAG, ESSL, IMSL). As in [51], we begin by developing sequential code for the radix 2 FFT, starting from a high–level specification, via a sequence of optimizing algebra–based transformations. This sequential code provides a common starting point for the development of sequential code for the general radix FFT in [51], and for the development of parallel/distributed algorithms presented here. For the convenience of the reader and to provide context for the transformations used here, we repeat this common development here in Section 3. Similarly, for the convenience of the reader, we also recall relevant experimental results from [51] in Section 7. These experiments provide evidence that the design methodology outlined here can already produce competitive code.

## 2. High–Level Description of Algorithm.

**2.1. Basic Algorithm.** Our approach to algorithm development begins by expressing the algorithm in a suitable high–level mechanism. This specification may take the form of a high–level statement of the algorithm, possibly annotated with additional specifications, such as the specification of the array of weights for the FFT. Our design and development of efficient implementations of an algorithm for the FFT began with Van Loan's [52] suitably commented high–level MATLAB program for the *radix 2* FFT, shown in Figure 2.1.

> **Input:** $x$ in $C^n$ and $n = 2^t$, where $t \geq 1$ is an integer.
> **Output:** The FFT of $x$.
>
> $$\begin{aligned}
> & x \leftarrow P_n \, x & & (1) \\
> & \text{for } q = 1 \text{ to } t & & (2) \\
> & \quad \text{begin} & & (3) \\
> & \quad\quad L \leftarrow 2^q & & (4) \\
> & \quad\quad r \leftarrow n/L & & (5) \\
> & \quad\quad x_{L \times r} \leftarrow B_L \, x_{L \times r} & & (6) \\
> & \quad \text{end} & & (7)
> \end{aligned}$$
>
> Here, $P_n$ is a $n \times n$ permutation matrix. Letting $L_* = L/2$, and $\omega_L$ be the $L$'th root of unity, matrix $B_L = \begin{bmatrix} I_{L_*} & \Omega_{L_*} \\ I_{L_*} & -\Omega_{L_*} \end{bmatrix}$, where $\Omega_{L_*}$ is a diagonal matrix with values $1, \omega_L, \ldots, \omega_L^{L_*-1}$ along the diagonal.

Fig. 2.1. *High–level algorithm for the radix 2 FFT*

In Line 1, $P_n$ is a permutation matrix that performs the bit–reversal permutation on the $n$ elements of vector $x$. The reference to $x_{L \times r}$ in Line 6 can be regarded as *reshaping* the $n$ element array $x$ to be a $L \times r$ matrix consisting of $r$ columns, where each column can be viewed as a vector of $L$ elements. This reshaping of $x$ is column–wise, so that each time Line 6 is executed, each pair of adjacent columns of the preceding matrix are concatenated to produce a single column of the new matrix. Line 6 can be viewed as treating each column of this matrix as a pair of vectors, each with $L/2$ elements, and doing a *butterfly* computation that combines the two $L/2$ element vectors in each column to produce a vector with $L$ elements. The butterfly computation, corresponding to multiplication of the data matrix $x$ by $B_L$, combines each pair of $L/2$ element column vectors from the old matrix into a new $L$ element vector for each column of the new matrix.

**2.2. Components of the Basic Algorithm.** By inspection of Figure 2.1, one can identify the following five components of the high–level *radix 2* FFT algorithm:

1. The computation of the bit–reversal permutation (Line 1).
2. The computation of the complex weights occurring in the matrices $\Omega_{L_*}$. These weights are discussed further in Section 3.1.
3. The *butterfly* computation that, using the weights, combines two vectors from $x$, producing a vector with twice as many elements (Line 6). The butterfly computation is scalarized, and subsequently refined in Sections 3.2–3.7. Contiguous memory access gives better performance at all levels of a memory hierarchy. Consequently, making data access during the butterfly computation contiguous is a driving factor throughout this paper in refining the butterfly computation.
4. The strategy for allocating the elements of the reshaped array $x$ (line 6) to the processors for use in parallel and distributed implementations of the computation loop (the $q$ loop of Lines 2 through 7). Alternative strategies for which processor will compute which values of $x$ during each iteration of the computation loop are discussed in Section 4. The actual location of the data is discussed in Sections 5 and 6.
5. The generation and bundling of messages involving values from $x$ (Lines 2 through 7). Message passing of data in distributed computation is discussed in Section 5.

**3. Development of Sequential Code.**

**3.1. Specification of the Matrix of Weight Constants.** The description of the algorithm in Figure 2.1 uses English to describe the constant matrix $B_L$. This matrix is a large two-dimensional sparse array. Of more importance for our purposes, this array can be naturally specified by using constructors from the array calculus with a few dense one dimensional vectors as the basis. This yields a succinct description of how $B_L$ can be constructed via a composition of array operations. Indeed, $B_L$ is never actually materialized as a dense $L \times L$ matrix. Rather, the succinct representation is used to guide the generation of code for the FFT, and the generated code only has multiplications corresponding to multiplication by a nonzero element of $B_L$.

On the top–level, $B_L$ is constructed by the appropriate concatenation of four submatrices. Only the construction of two of these submatrices, namely $I_{L_*}$ and $\Omega_{L_*}$, need be separately specified. Matrix $I_{L_*}$ occurs twice in the decomposition of $B_L$. Matrix $-\Omega_{L_*}$ can be obtained from matrix $\Omega_{L_*}$ by applying element–wise unary minus.

Each of the two submatrices to be constructed is a diagonal matrix. For each of these diagonal matrices, the values along the diagonal are successive powers of a given scalar. Psi Calculus contains a constructor that produces a vector whose elements are consecutive multiples or powers of a given scalar value. There is another constructor **diagonalize**[3] that converts a vector into a diagonal matrix with the values from the vector occurring along the diagonal, in the same order. We specified the matrices $I_{L_*}$ and $\Omega_{L_*}$ by using the vector constructor that produces successive powers, followed by the *diagonalize* constructor.

Since $L = 2^q$, matrix $B_L$ is different for each iteration of the loop in Figure 2.1. Accordingly, the specification of $I_{L_*}$ and $\Omega_{L_*}$ is *parameterized* by $L$ (and hence implicitly by $q$).

---
[3] The **diagonalize** operation is itself specified as a composition of more primitive array operations.

**3.2. Scalarization of the Matrix Multiplication.** A direct and easily automated scalarization of the matrix multiplication in Line 6 of Figure 2.1 produced code similar[4] to that given in Figure 3.1. Here `weight` is a vector of consecutive powers of $\omega_L$. Note that in Figure 3.1, the multiplication by the appropriate constant from the diagonal of matrix $\Omega_{L_*}$ is done explicitly, using a value from vector `weight`, and the multiplication by the constant 1 from the diagonal of matrix $I_{L_*}$ is done implicitly. Because the scalarized code does an assignment to one element at a time, the code stores the result of the array multiplication into a temporary array `xx`, and then copies `xx` into `x`. The sparseness of matrix $B_L$ is reflected in the assignments to `xx(row,col)`. The right–hand side of these assignment statements is the sum of only the two terms corresponding to the two nonzero elements of the row of $B_L$ involved in the computation of the left–hand side element. Moreover, the "regular" structure of $B_L$, expressed in terms of diagonal submatrices, provides a uniform way of selecting the two terms.

```
do col = 0,r-1
  do row = 0,L-1
    if (row < L/2 ) then
      xx(row,col) = x(row,col) + weight(row)*x(row+L/2,col)
    else
      xx(row,col) = x(row-L/2,col) - weight(row-L/2)*x(row,col)
    end if
  end do
end do
```

FIG. 3.1. *Direct Scalarization of the Matrix Multiplication*

**3.3. Representing Data as a 1–Dimensional Vector.** The data array $x$ in Figure 2.1 is a 2-dimensional array, that is reshaped during each iteration. Computationally, however, it is more efficient to store the data values in $x$ as a one dimensional vector, and to completely avoid performing this reshaping. Psi calculus easily handles avoiding array reshaping, and automates such transformations as the mapping of indices of an element of the 2–dimensional array into the index of the corresponding element of the vector. The $L \times r$ matrix $x_{L \times r}$ is stored as an $n$–element vector, which we denote as program variable `x`. The elements of the two–dimensional matrix are envisioned as being stored in *column–major* order, reflecting the column–wise reshaping occurring in Figure 2.1. Thus, element $x_{L \times r}(row, col)$ of $x_{L \times r}$ corresponds to element `x(L*col+row)` of `x`. Consequently, when $L$ changes, and matrix $x_{L \times r}$ is reshaped, no movement of data elements of vector `x` actually occurs. Replacing each access to an element of the two dimensional matrix $x_{L \times r}$ with an access to the corresponding element of the vector `x`, produces the code shown in Figure 3.2.

As an alternative to the scalarized code of Figure 3.2, the outer loop variable can iterate over the starting index of each column in vector `x`, using the appropriate stride to increment the loop variable. Instead of using a loop variable `col`, which ranges from 0 to `r-1` with a stride of 1, we use a variable, say `col'`, such that `col'` = L * col, and which consequently has a stride of L. By doing this, we eliminate the multiplication L * col that occurs each time an element of the arrays `xx` or `x` is accessed. This form of scalarization produces the code shown in Figure 3.3.

---

[4] The data array was not reshaped for each value of $q$, so references to the data array elements was more complicated than shown in Figure 3.1.

```
    do col = 0,r-1
      do row = 0,L-1
        if (row < L/2 ) then
          xx(L*col+row) = x(L*col+row) + weight(row)*x(L*col+row+L/2)
        else
          xx(L*col+row) = x(L*col+row-L/2) - weight(row-L/2)*x(L*col+row)
        end if
      end do
    end do
```

Fig. 3.2. *Scalarized Code with Data Stored in a Vector*

```
    do col' = 0,n-1,L
      do row = 0,L-1
        if (row < L/2 ) then
          xx(col'+row) = x(col'+row) + weight(row)*x(col'+row+L/2)
        else
          xx(col'+row) = x(col'+row-L/2) - weight(row-L/2)*x(col'+row)
        end if
      end do
    end do
```

Fig. 3.3. *Striding through the Data Vector*

**3.4. Elimination of Conditional Statement.** The use of the conditional statement that tests `row < L/2` in the above code can be eliminated automatically, as follows. We first *re–tile* the loop structure so that the innermost loop iterates over each pair of data elements that participate in each butterfly combination. To accomplish this re–tiling, we first envision reshaping the two–dimensional array $x_{L \times r}$ into a three–dimensional array $x_{L/2 \times 2 \times r}$. Under this reshaping, element $x_{L \times r}(row, col)$ corresponds to element $x_{L/2 \times 2 \times r}(row \bmod L/2, \lfloor row/(L/2) \rfloor, col)$. The middle dimension of $x_{L/2 \times 2 \times r}$ splits each column of $x_{L \times r}$ into the upper and lower parts of the column. Scalarizing Line 6 of Figure 2.1 based on the three–dimensional array $x_{L/2 \times 2 \times r}$, and indexing over the third dimension in the outer loop, over the first dimension in the middle loop, and over the second dimension in the innermost loop, produces the code shown in Figure 3.4.

To eliminate the conditional statement, we unroll the innermost loop, and produce the code shown in Figure 3.5.

When we represent the data array as a one–dimensional vector, the code shown in Figure 3.6 is produced.

**3.5. Optimizing the Basic Block.** The basic block in the inner loop of the above code has two occurrences of the common subexpression `weight(row)*x(col'+row+L/2)`. We hand–optimized this basic block, to compute this common subexpression only once. This produces more efficient code for the basic block, as shown in Figure 3.7.

Incorporating all the transformations described so far, the loop in Figure 2.1 is scalarized as shown in Figure 3.8. In this code, `pi` and `i` are Fortran "parameters", i.e., named constants.

**3.6. Doing the Butterfly Computation In–Place.** The use of the temporary array `xx` in the butterfly computation is unnecessary, and can be avoided by the

```
  do col = 0,r-1
    do row = 0,L/2-1
      do group = 0,1
        if ( group == 0 ) then
          xx(row,group,col) =
                  x(row,group,col) + weight(row)*x(row,group+1,col)
        else
          xx(row,group,col) =
                  x(row,group-1,col) - weight(row)*x(row,group,col)
        end if
      end do
    end do
  end do
```

Fig. 3.4. *Re–tiled Loop*

```
  do col = 0,r-1
    do row = 0,L/2-1
      xx(row,0,col) = x(row,0,col) + weight(row)*x(row,1,col)
      xx(row,1,col) = x(row,0,col) - weight(row)*x(row,1,col)
    end do
  end do
```

Fig. 3.5. *Unrolled Inner Loop*

use of a scalar variable to hold the value of `x(col'+row)`. Code incorporating this modification is shown in Figure 3.9, where scalar variable `d` is used for this purpose.

**3.7. Vectorizing the Butterfly Computation.** An alternative coding style for the butterfly computation is to use monolithic vector operations applied to appropriate sections of the data array. This vectorization of the butterfly computation produces the code shown in Figure 3.10[5]. Here, `cvec` is a one–dimensional array used to store the vector of values assigned to variable `c` during the iterations of the inner loop (the `row` loop) in Figure 3.9. Similarly, `dvec` is a one–dimensional array used to store the vector of values assigned to variable `d` during the iterations of this inner loop.

**4. Development of Plans for Parallel and Distributed Code.**

**4.1. An Overview of Plans, Templates, and Architectural Issues.** We want to generate code for two types of architecture.
- Distributed memory architecture (where each processor has its own local address space).
- Shared memory architecture with substantial amounts of local memory (where there is a common address space that includes the local memories).

To accommodate these architectures, we will develop an appropriate *parallel computation plan*, followed by appropriate *parallel computation templates*.

What we mean by a *parallel computation plan* is sequential code that indicates which parts of the overall work is to be done by each individual processor. A plan

---

[5] We assume that code can select a set of array elements using a start, stop, and stride mechanism, with a default stride of 1.

```
   do col' = 0,n-1,L
     do row = 0,L/2-1
       xx(col'+row) = x(col'+row) + weight(row)*x(col'+row+L/2)
       xx(col'+row+L/2) = x(col'+row) - weight(row)*x(col'+row+L/2)
     end do
   end do
```

Fig. 3.6. *Unrolled Inner Loop With Data Stored as a Vector*

```
   c = weight(row)*x(col'+row+L/2)
   xx(col'+row) = x(col'+row) + c
   xx(col'+row+L/2) = x(col'+row) - c
```

Fig. 3.7. *Optimized Basic Block*

specifies for each iteration of an outer loop[6], which iterations of inner loops should be performed by each of the processors of a multiprocessor system. At each point in the parallel computation, we envision that *responsibility* for the $n$ elements is partitioned among the processors. Our intention is that this responsibility is based on an *owner computes* rule, namely a given processor is responsible for those data elements for which it executes the butterfly assignment.

What we mean by a *template* is distributed or parallel "generic code" that indicates not only which parts of the overall work is to be done by each individual processor, but also where data is located, and how data is moved. To convert a template into code, a specific programming language needs to be selected, and more detail needs to be filled in. We use two styles of templates. The first style is a *per–processor* style, motivated by MPI [24]. The other style is an *all–processor* style, motivated by OpenMP [25].

It is our intention that in the transformation from an FFT parallel computation plan into a template for a distributed architecture or a shared memory architecture where substantial local memory is available to each processor, each of the processors will hold in its local memory those data elements it is responsible for, and do the butterfly operations on these data elements.

**4.2. Splitting the Outer Loop of the Butterfly Computation.** Suppose there are $m$ processors, where $m$ is a power of 2. We let *psize* denote $n/m$, the number of elements that each processor is responsible for. Our parallel computation plans will assume that $psize \geq m$. Now envision the data in the form of a two–dimensional matrix that gets reshaped for each value of $q$, as in Figure 2.1. During the initial rounds of the computation, for each processor, the value of *psize* is large enough to hold one or more columns of the two–dimensional data matrix, so we will make each processor responsible for multiple columns, where the total number of elements in these columns equals *psize*. For each iteration of $q$, the length of a column doubles, so that at some point, a column of the matrix will have more than *psize* elements. This first occurs when $q$ equals $\log_2(psize) + 1$. We call this value *breakpoint*. Once $q$ equals *breakpoint*, each column has more than *psize* elements. Consequently, we no longer want only one processor to be responsible for an entire column, and so we change plans. Before $q$ equals *breakpoint*, the number of columns is a multiple of $m$.

---

[6] For the FFT, this outer loop is the $q$ loop of Figure 3.9.

```
do q = 1,t
  L = 2**q
  do row = 0,L/2-1
    weight(row) = EXP((2*pi*i*row)/L)
  end do
  do col' = 0,n-1,L
    do row = 0,L/2-1
      c = weight(row)*x(col'+row+L/2)
      xx(col'+row) = x(col'+row) + c
      xx(col'+row+L/2) = x(col'+row) - c
    end do
  end do
  x = xx
end do
```

Fig. 3.8. *Loop with Optimized Basic Block*

```
do q = 1,t
  L = 2**q
  do row = 0,L/2-1
    weight(row) = EXP((2*pi*i*row)/L)
  end do
  do col' = 0,n-1,L
    do row = 0,L/2-1
      c = weight(row)*x(col'+row+L/2)
      d = x(col'+row)
      x(col'+row) = d + c
      x(col'+row+L/2) = d - c
    end do
  end do
end do
```

Fig. 3.9. *Loop with In–Place Butterfly Computation*

From *breakpoint* on, there are fewer than $m$ columns, but the number of rows is a multiple of $m$.

As long as $q$ is less than *breakpoint*, we can use a *block* approach to the computation, with each processor computing the new values of several *consecutive* columns of the two–dimensional matrix. Assume that the processors are numbered 0 through $m-1$. In terms of the one–dimensional vector of data elements, the columns whose new values are to be computed by processor $p$ are the *psize* consecutive vector elements beginning with vector element $p*psize$. When $q$ equals *breakpoint*, the number of elements in each column is 2 * *psize*, whereas we want each processor to compute the new value of only *psize* elements. Thus, when $q$ equals *breakpoint*, we need to switch to a different plan. To facilitate the switch to a different plan, we first modify Figure 3.9 by splitting the $q$ loop into two separate loops, as shown in Figure 4.1.

**4.3. Parallel Computation Plan When Local Memory is Available.** Consider the two $q$ loops in Figure 4.1. For the first $q$ loop, we will use a *block* approach to splitting the computation among the processors, as discussed in Section 4.2. Recall that before $q$ equals *breakpoint*, the number of columns is a multiple of $m$, and from

```
   do col' = 0,n-1,L
      cvec(0:L/2-1) = weight(0:L/2-1) * x(col'+L/2:col'+L-1)
      dvec(0:L/2-1) = x(col':col'+L/2-1)
      x(col':col'+L/2-1) = dvec(0:L/2-1) + cvec(0:L/2-1)
      x(col'+L/2:col'+L-1) = dvec(0:L/2-1) - cvec(0:L/2-1)
   end do
```

Fig. 3.10. *Vectorizing the Butterfly Computation*

```
do q = 1,breakpoint - 1
   L = 2**q
   do row = 0,L/2-1
      weight(row) = EXP((2*pi*i*row)/L)
   end do
   do col' = 0,n-1,L
      do row = 0,L/2-1
         c = weight(row)*x(col'+row+L/2)
         d = x(col'+row)
         x(col'+row) = d + c
         x(col'+row+L/2) = d - c
      end do
   end do
end do
do q = breakpoint,t
   L = 2**q
   do row = 0,L/2-1
      weight(row) = EXP((2*pi*i*row)/L)
   end do
   do col' = 0,n-1,L
      do row = 0,L/2-1
         c = weight(row)*x(col'+row+L/2)
         d = x(col'+row)
         x(col'+row) = d + c
         x(col'+row+L/2) = d - c
      end do
   end do
end do
```

Fig. 4.1. *Loop Splitting the Butterfly Computation*

*breakpoint* on, the number of rows is a multiple of $m$. For the second $q$ loop, which begins with $q$ equal to *breakpoint*, we use a *cyclic* approach, with each processor $p$ computing the new values of all the rows $j$ of the two–dimensional matrix such that $j$ is congruent to $p$ mod $m$. This corresponds to all the iterations of the inner loop where program variable row is congruent to $p$ mod $m$.

We express a computation plan by modifying each of the $q$ loops in Figure 4.1, so that for each value of $q$, there is a new outer loop using a new loop variable $p$, which ranges between 0 and $m - 1$. Our intention is that all the computations within the loop for a given value of $p$ will be performed by processor $p$. Figure 4.2 shows the effect of restructuring each of the two $q$ loops in Figure 4.1 by introducing a $p$ loop to reflect the parallel computation plan. The first $q$ loop uses a block approach for each $p$, and the second $q$ loop uses a cyclic approach for each $p$.

```
do q = 1,breakpoint - 1
  L = 2**q
  do row = 0,L/2-1
    weight(row) = EXP((2*pi*i*row)/L)
  end do
  do p = 0,m-1
    do col' = p*psize,(p+1)*psize-1,L
      do row = 0,L/2-1
        c = weight(row)*x(col'+row+L/2)
        d = x(col'+row)
        x(col'+row) = d + c
        x(col'+row+L/2) = d - c
      end do
    end do
  end do
end do
do q = breakpoint,t
  L = 2**q
  do row = 0,L/2-1
    weight(row) = EXP((2*pi*i*row)/L)
  end do
  do p = 0,m-1
    do col' = 0,n-1,L
      do row = p,L/2-1,m
        c = weight(row)*x(col'+row+L/2)
        d = x(col'+row)
        x(col'+row) = d + c
        x(col'+row+L/2) = d - c
      end do
    end do
  end do
end do
```

FIG. 4.2. *Initial Parallel Computation Plan*

Each execution of a $p$ loop in Figure 4.2 consists of $m$ iterations. Note that for each value of $q$, the sets of elements of the $x$ vector accessed by these $m$ iterations are pairwise disjoint. Our intention is that each of these $m$ iterations will be performed by a different processor. Since these processors will be accessing disjoint sets of elements from the $x$ vector, the execution of different iterations of a $p$ loop by different processors can be done in parallel.

Now consider the first $q$ loop in Figure 4.2. For each value of $p$ within this first $q$ loop, the data elements accessed are the *psize* elements beginning at position $p*psize$ of $x$. Since for a given $p$, these are the same set of data elements for every value of $q$ in the outer loop, we can interchange the loop variables $q$ and $p$ in the first $q$ loop. Now consider the second $q$ loop in Figure 4.2. For each value of $p$ within this second $q$ loop, the data elements accessed are the *psize* elements of $x$ whose position is congruent to $p$ mod $m$. Since for a given $p$, these are the same set of data elements for every value of $q$ in the outer loop, we can interchange the loop variables $q$ and $p$ in the second $q$ loop. This analysis shows that we can interchange the loop variables $q$ and $p$ in each of the two loops in Figure 4.2 to make $p$ the outermost loop variable, resulting in Figure 4.3. Note that a consequence of this loop interchange is that the computation of the *weight* vector for each value of $q$ is now done for each value of $p$.

```
do p = 0,m-1
  do q = 1,breakpoint - 1
    L = 2**q
    do row = 0,L/2-1
      weight(row) = EXP((2*pi*i*row)/L)
    end do
    do col' = p*psize,(p+1)*psize-1,L
      do row = 0,L/2-1
        c = weight(row)*x(col'+row+L/2)
        d = x(col'+row)
        x(col'+row) = d + c
        x(col'+row+L/2) = d - c
      end do
    end do
  end do
end do
do p = 0,m-1
  do q = breakpoint,t
    L = 2**q
    do row = 0,L/2-1
      weight(row) = EXP((2*pi*i*row)/L)
    end do
    do col' = 0,n-1,L
      do row = p,L/2-1,m
        c = weight(row)*x(col'+row+L/2)
        d = x(col'+row)
        x(col'+row) = d + c
        x(col'+row+L/2) = d - c
      end do
    end do
  end do
end do
```

FIG. 4.3. *Local Memory Parallel Computation Plan with Processor Loops Outermost*

**4.4. Computing Only Needed Weights.** In the second $q$ loop in Figure 4.3, each $p$ computes an entire *weight* vector of $L/2$ elements. However, for a given $p$, only those elements in the *weight* vector whose position is congruent to $p$ mod $m$ are actually used. We can change the plan so that only these $\frac{L}{2m}$ elements of the *weight* vector are computed. The computation of weights in the second $q$ loop would then be as follows.

```
      do row = p,L/2-1,m
         weight(row) = EXP((2*pi*i*row)/L)
      end do
```

**4.5. Making Weight Vector Access Contiguous.** Note that in the second $q$ loop, the *weight* vector is accessed with consecutive values of the program variable *row*. These consecutive values of *row* are strided, with a stride of $m$. Consequently, the accesses of the *weight* vector in the second $q$ loop are strided, with a stride of $m$. Within each iteration of the second $q$ loop we can make the accesses to the *weight* vector contiguous, as follows. The key is to use a new weight vector `weightcyclic`, in which we store only the $\frac{L}{2m}$ weight elements needed for a given value of $p$. Consequently, these needed elements of *weightcyclic* will be accessed contiguously. The relationship between the *weightcyclic* and *weight* vectors is that for each $j$, $0 \leq j < \frac{L}{2m}$, $weightcyclic(j)$ will hold the value of $weight(m*j+p)$, i.e., for each value of *row* occurring in an inner loop with loop control

```
      do row = p,L/2-1,m
```

the value of $weight(row)$ will be held in $weightcyclic(\frac{row-p}{m})$.

With this change, the computation of weights in the second $q$ loop becomes the following

```
      do row = p,L/2-1,m
         weightcyclic((row-p)/m) = EXP((2*pi*i*row)/L)
      end do
```

We can simplify the loop control and subscript computation for *weightcyclic* by changing the loop control variable. In the loop which computes the needed weight elements, we use a loop index variable $row'$, where $row' = (row-p)/m$, i.e. $row = m*row'+p$.

With this change, the weight computation within the second $q$ loop becomes the following, where $row'$ has a stride of 1.

```
      do row' = 0,L/(2*m)-1,1
         weightcyclic(row') = EXP((2*pi*i*(m*row'+p))/L)
      end do
```

Within the butterfly computation, where the loop variable is *row*, a use of $weight(row)$ becomes a use of $weightcyclic((row-p)/m)$.

Figure 4.4 shows the second $q$ loop after incorporating the improvements from Section 4.4 and this Section to the computation and access of weights.

**4.6. Making Data Access Contiguous.** Note that the accesses of data vector $x$ in the second $q$ loop are strided, with a stride of $m$. We can make access to the needed data in $x$ in the second $q$ loop contiguous, as follows. We introduce a new data vector `xcyclic` to hold *only* those data elements accessed for each $p$. For each $p$, the vector *xcyclic* contains $L/m$ elements from each of the $n/L$ columns, for a total of $n/m = psize$ elements. The relationship between the *xcyclic* and $x$ vectors is that for each $col'$, where $0 \leq col' < n-1$ and $col'$ is a multiple of $L$, and $row$, where $0 \leq row < L$ and $row$ is congruent to $p$ mod $m$, $xcyclic(col'/m + (row-p)/m)$ will hold the value of $x(col'+row)$. Consequently, for each value of $col'$ and $row$ occurring in the butterfly loop, $x(col'+row)$ will be held in $xcyclic(col'/m + (row-p)/m)$ and $x(col'+row+L/2)$ will be held in $xcyclic(col'/m + (row+L/2-p)/m)$.

At the beginning of each iteration of the $p$ loop, the required *psize* elements of $x$ must be copied into *xcyclic*, and at the end of each iteration of the $p$ loop, these *psize* elements must be copied from *xcyclic* into $x$. At the template level, this copying will be done by either message passing or assignment statements, depending on the

```
      do p = 0,m-1
        do q = breakpoint,t
          L = 2**q
          do row' = 0,L/(2*m)-1,1
            weightcyclic(row') = EXP((2*pi*i*(m*row'+p))/L)
          end do
          do col' = 0,n-1,L
            do row = p,L/2-1,m
              c = weightcyclic((row-p)/m)*x(col'+row+L/2)
              d = x(col'+row)
              x(col'+row) = d + c
              x(col'+row+L/2) = d - c
            end do
          end do
        end do
      end do
```

FIG. 4.4. *Second q Loop with Contiguous Access of Needed Weights*

architecture and programming language. Here, where we are developing a compilation plan, we indicate this copying abstractly, using function calls. We envision array $x$ as containing a "centralized" copy of *all* the data elements, and *xcyclic* as containing a subset. The respective subsets of $x$ used by the iterations of the $p$ loop represent a *partition* of the data elements in $x$. We envision the $n$ elements of $x$ as being partitioned into $m$ subsets, where each iteration of the $p$ loop copies one of these subsets into *xcyclic*. The copying from $x$ to *xcyclic* of the *psize* elements of $x$ whose position in $x$ is congruent to $p$ mod $m$ is expressed using the function call

CENTRALIZED_TO_CYCLIC_PARTITIONED(x,xcyclic,m,psize,p).

The copying of the *psize* elements of *xcyclic* into $x$ is expressed using the function call

CYCLIC_PARTITIONED_TO_CENTRALIZED(xcyclic,x,m,psize,p).

Figure 4.5 shows the effect of incorporating these changes to the second $q$ loop. Note that since the stride in the *row* loop is $m$, these changes do indeed provide contiguous data access.

We can simplify the loop control and subscript computation of vector elements occurring in Figure 4.5, by changing the loop control variables, as shown in Figure 4.6.

In Figure 4.6, we use a variable `col''` to range over all the columns, where $col' = m * col''$ i.e. $col'' = col'/m$. Since the bounds for loop control variable `col'` in Figure 4.5 are 0,n-1,L, the bounds for the loop control variable `col''` in Figure 4.6 are 0,psize-1,L/m.

Similarly, in Figure 4.6, we use a variable `row'`, where $row = m * row' + p$, i.e. $row' = (row - p)/m$. Consider the bounds of the `row` loop in the second $q$ loop; the start, stop, stride are p,L/2-1,m. For the corresponding `row'` loop, the start becomes 0, and the stop becomes $\frac{L}{2m} - \frac{p+1}{m}$. Since $0 \le p < m$, we can conclude that $0 < \frac{p+1}{m} \le 1$, so the stop can be simplified to $\frac{L}{2m} - 1$. Finally, the step is 1. So, the bounds for the `row'` loop become 0,L/(2*m)-1,1.

Consequently, `xcyclic(col''+row')` will contain the value of `x(m*col''+m*row'+p)`.

**4.7. Facilitating Exploitation of Local Memory.** Note that in Figure 4.6, there is no data flow involving `xcyclic` between the iterations of the $p$ loop. A consequence is that when the plan is subsequently transformed into a template for

```
  do p = 0,m-1
    CENTRALIZED_TO_CYCLIC_PARTITIONED(x,xcyclic,m,psize,p)
    do q = breakpoint,t
      L = 2**q
      do row' = 0,L/(2*m)-1,1
        weightcyclic(row') = EXP((2*pi*i*(m*row'+p))/L)
      end do
      do col' = 0,n-1,L
        do row = p,L/2-1,m
          c = weightcyclic((row-p)/m) *xcyclic(col'/m+(row-p)/m+L/(2*m))
          d = xcyclic(col'/m+(row-p)/m)
          xcyclic(col'/m+(row-p)/m) = d + c
          xcyclic(col'/m+(row-p)/m+L/(2*m)) = d - c
        end do
      end do
    end do
    CYCLIC_PARTITIONED_TO_CENTRALIZED(xcyclic,x,m,psize,p)
  end do
```

FIG. 4.5. *Plan for Second q Loop with Contiguous Data Access*

```
  do p = 0,m-1
    CENTRALIZED_TO_CYCLIC_PARTITIONED(x,xcyclic,m,psize,p)
    do q = breakpoint,t
      L = 2**q
      do row' = p,L/2-1,m
        weightcyclic(row') = EXP((2*pi*i*(m*row'+p))/L)
      end do
      do col'' = 0,psize-1,L/m
        do row' = 0,L/(2*m)-1,1
          c = weightcyclic(row')*xcyclic(col''+row'+L/(2*m))
          d = xcyclic(col''+row')
          xcyclic(col''+row') = d + c
          xcyclic(col''+row'+L/(2*m)) = d - c
        end do
      end do
    end do
    CYCLIC_PARTITIONED_TO_CENTRALIZED(xcyclic,x,m,psize,p)
  end do
```

FIG. 4.6. *Plan for Second q Loop with Contiguous Data Access and Simplified Loop Control*

parallel and distributed code, xcyclic can be a local variable of each processor.

A similar transformation to that in Section 4.6 can be done to the second $q$ loop in the plan, although the payoff only occurs subsequently when templates are constructed from the plan. We can facilitate making access to the data in the first $q$ loop be via a processor–local variable. This can be accomplished by introducing a new data vector xblock, to hold the data values accessed by each iteration of the $p$ loop.

For each $p$, we will use the vector xblock to hold the *psize* elements accessed during that iteration of the $p$ loop. The relationship between the *xblock* and $x$ vectors is that for each $col'$, where $0 \leq col' < n-1$ and $col'$ is a multiple of $L$, and $row$, where $0 \leq row < L$, $x(col' + row)$ will be held in $xblock(col' + row - p*psize)$.

Figure 4.7 shows the effect of incorporating the use of *xblock* in the first $q$ loop.

```
do p = 0,m-1
  CENTRALIZED_TO_BLOCK_PARTITIONED(x,xblock,m,psize,p)
  do q = 1,breakpoint - 1
    L = 2**q
    do row = 0,L/2-1
      weight(row) = EXP((2*pi*i*row)/L)
    end do
    do col' = p*psize,(p+1)*psize-1,L
      do row = 0,L/2-1
        c = weight(row)*xblock(col'+row+L/2-p*psize)
        d = xblock(col'+row-p*psize)
        xblock(col'+row-p*psize) = d + c
        xblock(col'+row+L/2-p*psize) = d - c
      end do
    end do
  end do
  BLOCK_PARTITIONED_TO_CENTRALIZED(xblock,x,m,psize,p)
end do
```

FIG. 4.7. *Plan for First q Loop*

We can simplify the loop control and subscript computation of vector elements by changing the loop control variable $col'$, as shown in Figure 4.8. We use a variable `col''`, where $col'' = col' - p*psize$. So, the bounds for the loop control variable `col''` are 0,psize-1,L. Consequently, `xblock(col''+row)` will contain the value of `x(col'+row-p*psize)`.

Figure 4.9 shows the final plan, combining Figures 4.6 and 4.8.

**4.8. Flexibility in Setting Breakpoint.** Note that the split of the $q$ loop of Figure 3.9 into two separate $q$ loops, as shown in Figure 4.1, is correct for any value of *breakpoint*. However, for the transformation into the plan shown in Figure 4.2 to be correct, *breakpoint* must satisfy certain requirements, as follows.

Requirement 1: Consider the first $q$ loop in Figure 4.2. The correctness of this $q$ loop requires that $L \leq psize$, i.e. $2^q \leq psize$. Thus, each value of $q$ for which the loop is executed must satisfy $q \leq \log_2(psize)$. Since the highest value of $q$ for which the loop is executed is $breakpoint-1$, the first $q$ loop requires that $breakpoint \leq \log_2(psize)+1$.

Requirement 2: Consider the second $q$ loop in Figure 4.2. The correctness of this $q$ loop requires that $L/2 \geq m$, i.e. $2^{q-1} \geq m$. Thus, each value of $q$ for which the loop is executed must satisfy $q \geq \log_2(m)+1$. Since the lowest value of $q$ for which the loop is executed is *breakpoint*, the second $q$ loop requires that $breakpoint \geq \log_2(m) + 1$.

Together, these two requirements imply that:

$$\log_2(m) + 1 \leq breakpoint \leq \log_2(psize) + 1.$$

Since we are assuming that $m \leq psize$, there is a range of possible allowable values for *breakpoint*. Thus far in Section 4, we have assumed that *breakpoint* is set

```
      do p = 0,m-1
        CENTRALIZED_TO_BLOCK_PARTITIONED(x,xblock,m,psize,p)
        do q = 1,breakpoint - 1
          L = 2**q
          do row = 0,L/2-1
            weight(row) = EXP((2*pi*i*row)/L)
          end do
          do col'' = 0,psize-1,L
            do row = 0,L/2-1
              c = weight(row)*xblock(col''+row+L/2)
              d = xblock(col''+row)
              xblock(col''+row) = d + c
              xblock(col''+row+L/2) = d - c
            end do
          end do
        end do
        BLOCK_PARTITIONED_TO_CENTRALIZED(xblock,x,m,psize,p)
      end do
```

FIG. 4.8. *Plan for First q Loop with Simplified Loop Control*

at the high end of its allowable range, namely at $\log_2(psize) + 1$. In fact, *breakpoint* could be set anywhere within its allowable range. If the plan shown in Figure 4.4 is used, there is an advantage in setting *breakpoint* to the low end of its allowable range, for the following reason. Moving some values of $q$ from the first $q$ loop to the second $q$ loop would reduce the number of weights that need to be computed by each processor.

**5. Development of Per–Processor Templates for Distributed Memory Architectures.**

**5.1. Distributed Template Using Local Copy of Data Vector.**

**5.1.1. Partial Per–Processor Distributed Code Template.** Figure 5.1 shows how the parallel computation plan from Figure 4.3 can be transformed into a partially specified template (subsequently abbreviated as a *partial template*) for distributed code[7]. The template in Figure 5.1 is intended to represent per–processor code to be executed on each of the $m$ processors. We assume that each processor has a variable, named `myid`, whose value is the processor number, where the processor number ranges from 0 to $m - 1$. Thus, each of the $m$ processors, say processor $i$, has its value of local variable `myid` equal to $i$. In Figure 4.3, each $p$ loop has $m$ iterations, with $p$ ranging from 0 to $m - 1$. In Figure 5.1, variable `myid` is used to ensure that processor $i$ executes iteration $i$ of each $p$ loop, i.e., the iteration where $p$ equals $i$.

A `SYNCHRONIZE` command is inserted between the two $q$ loops to indicate that some form of synchronization between processors is needed because data computed by each processor in its first $q$ loop is subsequently accessed by all the processors in their second $q$ loop. We call Figure 5.1 a *partial* template because it does not address the issue of data location or data movement between processors. These issues must be resolved, and the `SYNCHRONIZE` command between the two $q$ loops must be replaced

---

[7] Any of the improvements from Section 4, such as computing only the needed weights, as discussed in Section 4.4, could be incorporated into the plan prior to this transformation, but for simplicity, here we give a transformation from the plan in Figure 4.3.

```
   do p = 0,m-1
     CENTRALIZED_TO_BLOCK_PARTITIONED(x,xblock,m,psize,p)
     do q = 1,breakpoint - 1
       L = 2**q
       do row = 0,L/2-1
         weight(row) = EXP((2*pi*i*row)/L)
       end do
       do col″ = 0,psize-1,L
         do row = 0,L/2-1
           c = weight(row)*xblock(col″+row+L/2)
           d = xblock(col″+row)
           xblock(col″+row) = d + c
           xblock(col″+row+L/2) = d - c
         end do
       end do
     end do
     BLOCK_PARTITIONED_TO_CENTRALIZED(xblock,x,m,psize,p)
   end do
   do p = 0,m-1
     CENTRALIZED_TO_CYCLIC_PARTITIONED(x,xcyclic,m,psize,p)
     do q = breakpoint,t
       L = 2**q
       do row′ = p,L/2-1,m
         weightcyclic(row′) = EXP((2*pi*i*(m*row′+p))/L)
       end do
       do col″ = 0,psize-1,L/m
         do row′ = 0,L/(2*m)-1,1
           c = weightcyclic(row′)*xcyclic(col″+row′+L/(2*m))
           d = xcyclic(col″+row′)
           xcyclic(col″+row′) = d + c
           xcyclic(col″+row′+L/(2*m)) = d - c
         end do
       end do
     end do
     CYCLIC_PARTITIONED_TO_CENTRALIZED(xcyclic,x,p,m,psize)
   end do
```

FIG. 4.9. *Combined Plan Using Partitioned Data*

by statements using an appropriate implementation language mechanism that permits each processor to proceed past the synchronization point only after all the processors have reached the synchronization point. These statements must ensure that the data needed by each processor is indeed available to that processor.

**5.1.2. Responsibility–Based Distribution of Data Vector on Local Memories.** In order to transform the partial template of Figure 5.1 into a (fully specified) template, we need to address the issue of where (i.e., on which processor) data is located, and how data moves between processors. A straightforward approach is to let each processor have space for an entire data vector x of $n$ elements, but in each of the two $q$ loops, only access those *psize* elements of x that the processor is responsible

```
      do q = 1,breakpoint - 1
        L = 2**q
        do row = 0,L/2-1
          weight(row) = EXP((2*pi*i*row)/L)
        end do
        do col' = myid*psize,(myid+1)*psize-1,L
          do row = 0,L/2-1
            c = weight(row)*x(col'+row+L/2)
            d = x(col'+row)
            x(col'+row) = d + c
            x(col'+row+L/2) = d - c
          end do
        end do
      end do
      SYNCHRONIZE
      do q = breakpoint,t
        L = 2**q
        do row = 0,L/2-1
          weight(row) = EXP((2*pi*i*row)/L)
        end do
        do col' = 0,n-1,L
          do row = myid,L/2-1,m
            c = weight(row)*x(col'+row+L/2)
            d = x(col'+row)
            x(col'+row) = d + c
            x(col'+row+L/2) = d - c
          end do
        end do
      end do
```

FIG. 5.1. *Partial Per–Processor Parallel Code Template Obtained from Plan of Figure 4.3*

for. To emphasize that space for all $n$ elements of a vector x is allocated on each processor, we refer to this space on a given processor $p$ as $x_p$.

For a distributed architecture or shared memory architecture with substantial local memories, the "current" value of each data element will reside on the processor that computes its new value, i.e., is responsible for it. We now consider this assignment of each element from the data array to the processor that is responsible for it. To start, consider the first $q$ loop in Figure 5.1. Since $q$ ranges between 1 and $breakpoint-1$, with $breakpoint= \log_2(psize)+1$ and $L = 2^q$, we can conclude that $L$ ranges between 2 and $psize$. For each processor $p$, consider the data elements accessed in each iteration of this first $q$ loop. These are data elements $x(col' + row)$ and $x(col' + row + L/2)$, where $col'$ ranges between $p*psize$ and $(p+1)*psize - 1$ in steps of $L$, and $row$ ranges between 0 and $L/2 - 1$. Since $L \leq psize$ for each iteration of the $q$ loop, the accessed data elements are $psize$ consecutive elements, beginning with element $x(p*psize)$. These elements can be described as $x(p*psize : ((p+1)*psize) - 1 : 1)$. This is a *block* distribution of the responsibility for array x, corresponding to the block approach to the first $p$ loop, as described in Section 4.3.

Now consider the second $q$ loop in Figure 5.1. In each iteration of this $q$ loop,

$L/2 \geq psize$. Since $psize \geq m$, we can conclude that $L/2 \geq m$. For each processor $p$, consider the data elements accessed in each iteration of the $q$ loop. These are data elements $x(col' + row)$ and $x(col' + row + L/2)$, where $col'$ ranges between 0 and $n-1$, in steps of $L$, and $row$ ranges between $p$ and $L/2 - 1$, in steps of $m$. Since $L/2$ is a multiple of $m$, so is $L$, so that $col' \equiv 0 \bmod m$ for every value of $col'$. Also, $row \equiv p \bmod m$. Consequently, $col' + row$ and $col' + row + L/2$ are each congruent to $p \bmod m$. Thus, every data element accessed by processor $p$ is congruent to $p \bmod m$. Now, note that there are $n/L$ iterations of the $col'$ loop. Because $L/2 \geq m$, for each $p$ and $col'$, there are $L/(2m)$ iterations of the $row$ loop. Each of these iterations does the butterfly operation on two data elements of array $x$. Note that

$$\frac{n}{L} * \frac{L}{2m} * 2 = \frac{n}{m} = psize.$$

Thus for each iteration of the $q$ loop, processor $p$ does an assignment to $psize$ data elements. We now note that these are distinct elements, since for each value of $row$ and $col'$, different data elements are accessed. In conclusion, the elements accessed by processor $p$ during each iteration of the loop are the set of elements $x(i)$ such that $i \equiv p \bmod m$. Using Fortran90–like start–stop–stride notation these elements can be described as $x(p : n-1 : m)$. This is a *cyclic* distribution of the responsibility for array x, corresponding to the cyclic approach to the second $p$ loop, as described in Section 4.3.

**5.1.3. Construction of Messages.** We now consider the construction of messages between processors for distributed computation. This entails both bundling of data values into messages, and having both the sender and recipient of each message specify which of its data elements are included in each message.

The data can be initially stored in various ways, such as on disks, distributed among the processors, or all on one processor. Each of these cases can be readily handled. Here, we envision that the data is initially centralized on one processor.

In describing the details of message passing, we assume that the numbering of processors from 0 to $m-1$ is such that initially all the data is on processor 0. Before the first $q$ loop, as per the block approach, the initially centralized data must be distributed among the processors. The processor with all the data must send every other processor, *otherp*, the set of *psize* contiguous data elements, beginning with data element $x(psize * otherp)$, and each such processor must receive these data elements. The recipient processor then stores the received *psize* data values in the appropriate position in its $x$ vector.

A high–level description of message–passing code to perform the block distribution is shown in Figure 5.2. Since this code refers to the variable `myid`, the same code can be used on each processor. Processor 0 will execute the *otherp* loop, sending a message to each of the other processors. Each processor other than processor 0 will execute a `RECEIVE` command. The `SEND` command sends a message, and has three parameters: the first element to be sent, the number of elements to send, and the processor which should receive the message. The `RECEIVE` command receives a message, and has three parameters: where to place the first element received, the number of elements to receive, and the processor which sent the message to be received. In Figure 5.2, we refer to the local data array as $x_{myid}$, to emphasize that it is the local space for the data array that is the source and destination of data sent and received by a given processor.

We assume that the `SEND` command is *nonblocking*, and uses *buffering*. Thus, the contents of the message to be sent is copied into a buffer, at which point the statement after the `SEND` command can be executed. We assume that the `RECEIVE` command is *blocking*. Thus, the data is received and stored in the locations specified by the command, at which point the statement after the `RECEIVE` command can be executed.

```
!  DISTRIBUTE_CENTRALIZED_TO_BLOCK(x_myid,m,psize)
if myid = 0 then
  do otherp = 1,m-1
    SEND(x_myid(psize*otherp),psize,otherp)
  end do
else
    RECEIVE(x_myid(psize*myid),psize,0)
endif
```

FIG. 5.2. *Centralized to Block Distribution*

Now consider the block to cyclic redistribution that occurs after the first $q$ loop, and before the second $q$ loop. Recall that each processor has *psize* data elements. Of the *psize* elements on any given processor before the redistribution, $psize/m$ must wind up in each of the processors (including itself) after the redistribution. No message need be sent for the data values that are to wind up on the same processor as before the redistribution. Thus, each processor must send $psize/m$ elements to each of the other processors. A high–level description of message–passing code to perform the block to cyclic redistribution is shown in Figure 5.3. We assume that the `SEND` and `RECEIVE` commands are flexible enough so that the first parameter can specify a set of array elements using notation specifying start, stop, and stride. The `SEND` commands can be *nonblocking* because for any given processor, disjoint sets of data elements are involved in any pair of message commands on that processor. An alternative to Figure 5.3 is to first do all the `SEND`s, and then do all the `RECEIVE`s.

```
!  DISTRIBUTE_BLOCK_TO_CYCLIC(x_myid,m,psize)
do otherp = 0,m-1
  if otherp ≠ myid then
    SEND(x_myid(psize*myid+otherp:psize*(myid+1)-1:m),psize/m,otherp)
    RECEIVE(x_myid(psize*otherp+myid:psize*(otherp+1)-1:m),psize/m,otherp)
  endif
end do
```

FIG. 5.3. *Block to Cyclic Redistribution*

Another alternative to Figure 5.3 is to first collect all the data at the centralized site, i.e. processor 0, and then do a centralized to cyclic redistribution. This strategy is described in more detail in Figure 5.4, where the block to cyclic redistribution is done as block to centralized redistribution, followed by a centralized to cyclic redistribution. The approach in Figure 5.4 entails only $2(m-1)$ messages, in contrast to the $m(m-1)$ messages entailed in Figure 5.3. However, the centralized site may become a bottleneck. Moreover, data elements are transmitted twice, to and from the centralized site, rather than only once.

We envision that at the end of the FFT computation all the data elements are collected at the centralized processor, processor 0. A high–level description of message–passing code to perform this cyclic to centralized redistribution is shown in Figure 5.5.

```
!   DISTRIBUTE_BLOCK_TO_CYCLIC(x_myid,m,psize)
DISTRIBUTE_BLOCK_TO_CENTRALIZED(x_myid,m,psize)
DISTRIBUTE_CENTRALIZED_TO_CYCLIC(x_myid,m,psize,n)

!   DISTRIBUTE_BLOCK_TO_CENTRALIZED(x_myid,m,psize)
if myid = 0 then
  do otherp = 1,m-1
    RECEIVE(x_myid(psize*otherp),psize,otherp)
  end do
else
    SEND(x_myid(psize*myid),psize,0)
endif

!   DISTRIBUTE_CENTRALIZED_TO_CYCLIC(x_myid,m,psize,n)
if myid = 0 then
  do otherp = 1,m-1
    SEND(x_myid(otherp:n-1:m),psize,otherp)
  end do
else
    RECEIVE(x_myid(myid:n-1:m),psize,0)
endif
```

FIG. 5.4. *Block to Cyclic Redistribution Using Centralized Site as Intermediary*

```
!   DISTRIBUTE_CYCLIC_TO_CENTRALIZED(x_myid,m,psize,n)
if myid = 0 then
  do otherp = 1,m-1
    RECEIVE(x_myid(otherp:n-1:m),psize,otherp)
  end do
else
    SEND(x_myid(myid:n-1:m),psize,0)
endif
```

FIG. 5.5. *Cyclic to Centralized Redistribution*

**5.1.4. Per–Processor Distributed Code Template.** Figure 5.6 shows how the partial parallel code template from Figure 5.1 is transformed into a template that specifies where data is located and how it moves between processors. In Figure 5.6, we refer to the arrays involved in the template as $\mathtt{x}_{myid}$ and $\mathtt{weight}_{myid}$, to emphasize that space local to a given processor is being accessed by the per–processor template. We assume that scalar variables, such as q and L are understood to be local variables within each processor. As discussed in Section 5.1.2, the distributed code template will perform a data redistribution from the initial centralized distribution to a block distribution, followed by the first $q$ loop, followed by a block to cyclic redistribution, followed by the second $q$ loop (which begins with $q$ equal to *breakpoint*), followed by a cyclic to centralized redistribution. The code in Figure 5.6 is executed on each of the $m$ processors. The DISTRIBUTE_BLOCK_TO_CYCLIC command in Figure 5.6 serves as the mechanism implementing the SYNCHRONIZE command in Figure 5.1. The block to cyclic redistribution can be performed either as described in Figure 5.3 or as described in Figure 5.4. In either case, the blocking RECEIVEs do the synchronization.

```
    DISTRIBUTE_CENTRALIZED_TO_BLOCK(x_myid,m,psize)
  do q = 1,breakpoint - 1
    L = 2**q
    do row = 0,L/2-1
      weight_myid(row) = EXP((2*pi*i*row)/L)
    end do
    do col' = myid*psize,(myid+1)*psize-1,L
      do row = 0,L/2-1
        c = weight_myid(row)*x_myid(col'+row+L/2)
        d = x_myid(col'+row)
        x_myid(col'+row) = d + c
        x_myid(col'+row+L/2) = d - c
      end do
    end do
  end do
  DISTRIBUTE_BLOCK_TO_CYCLIC(x_myid,m,psize)
  do q = breakpoint,t
    L = 2**q
    do row = 0,L/2-1
      weight_myid(row) = EXP((2*pi*i*row)/L)
    end do
    do col' = 0,n-1,L
      do row = myid,L/2-1,m
        c = weight_myid(row)*x_myid(col'+row+L/2)
        d = x_myid(col'+row)
        x_myid(col'+row) = d + c
        x_myid(col'+row+L/2) = d - c
      end do
    end do
  end do
  DISTRIBUTE_CYCLIC_TO_CENTRALIZED(x_myid,m,psize,n)
```

FIG. 5.6. *Per–Processor Distributed Code Template Based on Partial Template of Figure 5.1*

**5.2. Distributed Template with Partitioned Data and Contiguous Access.**

**5.2.1. Development of Template from Plan Using Partitioned Data.** Figure 5.7 shows a per–processor template obtained from Figure 4.9's parallel combined computation plan using partitioned data. The template in Figure 5.7 assumes that all $n$ data elements are initially located on processor 0 and stored in array $x_0$. It also assumes that at the end of the algorithm, all $n$ data elements are to be collected together on processor 0 and stored in array $x_0$. Each processor $p$ is assumed to have its own local arrays $\text{weight}_p$, $\text{xblock}_p$, $\text{weightcyclic}_p$, and $\text{xcyclic}_p$. The iteration over the $p$ loops in Figure 4.9 is replaced by the execution of the per–processor template in Figure 5.7 on each of the $m$ processors.

A high–level description of message–passing code to perform the four data redistribution commands in Figure 5.7 (centralized–to–block–partitioned, block–partitioned–to–centralized, centralized–to–cyclic–partitioned, and cyclic–partitioned–to–centralized) is shown in Figure 5.8.

```
   DISTRIBUTE_CENTRALIZED_TO_BLOCK_PARTITIONED(x_0,xblock_myid,m,psize)
do q = 1,breakpoint - 1
  L = 2**q
  do row = 0,L/2-1
    weight_myid(row) = EXP((2*pi*i*row)/L)
  end do
  do col'' = 0,psize-1,L
    do row = 0,L/2-1
      c = weight_myid(row)*xblock_myid(col''+row+L/2)
      d = xblock_myid(col''+row)
      xblock_myid(col''+row) = d + c
      xblock_myid(col''+row+L/2) = d - c
    end do
  end do
end do
DISTRIBUTE_BLOCK_PARTITIONED_TO_CENTRALIZED(xblock_myid,x_0,m,psize)
DISTRIBUTE_CENTRALIZED_TO_CYCLIC_PARTITIONED(x_0,xcyclic_myid,m,psize)
do q = breakpoint,t
  L = 2**q
  do row' = 0,L/(2*m)-1,1
    weightcyclic_myid(row') = EXP((2*pi*i*(m*row'+myid))/L)
  end do
  do col'' = 0,psize-1,L/m
    do row' = 0,L/(2*m)-1,1
      c = weightcyclic_myid(row')*xcyclic_myid(col''+row'+L/(2*m))
      d = xcyclic_myid(col''+row')
      xcyclic_myid(col''+row') = d + c
      xcyclic_myid(col''+row'+L/(2*m)) = d - c
    end do
  end do
end do
DISTRIBUTE_CYCLIC_PARTITIONED_TO_CENTRALIZED(xcyclic_myid,x_0,m,psize)
```

FIG. 5.7. *Per–Processor Distributed Code Template Obtained from Combined Plan Using Partitioned Data of Figure 4.9*

**5.2.2. Eliminating Bottleneck in Block to Cyclic Redistribution.** In Figure 5.7, the pair of data redistribution commands

DISTRIBUTE_BLOCK_PARTITIONED_TO_CENTRALIZED(xblock$_{myid}$,x$_0$,m,psize)
DISTRIBUTE_CENTRALIZED_TO_CYCLIC_PARTITIONED(x$_0$,xcyclic$_{myid}$,m,psize)

occurring between the two $q$ loops, together accomplish a block–partitioned–to–cyclic-partitioned data redistribution. This two–command sequence can be replaced by the single command

DISTRIBUTE_BLOCK_PARTITIONED_TO_CYCLIC_PARTITIONED(xblock$_{myid}$,xcyclic$_{myid}$,m,psize)

This combined command can be implemented by doing a block–partitioned–to–centralized redistribution followed by a centralized–to–block–partitioned redistribution, as indicated in Figure 5.7, in a manner analogous to that shown in Figure 5.4. Alternately, the block–partitioned–to–centralized redistribution can be done directly, as shown in

```
!   DISTRIBUTE_CENTRALIZED_TO_BLOCK_PARTITIONED(x_0,xblock_{myid},m,psize)
if myid = 0 then
  xblock_{myid} = x_0(0:psize-1)
  do otherp = 1,m-1
    SEND(x_0(psize*otherp),psize,otherp)
  end do
else
    RECEIVE(xblock_{myid},psize,0)
endif

!   DISTRIBUTE_BLOCK_PARTITIONED_TO_CENTRALIZED(xblock_{myid},x_0,m,psize)
if myid = 0 then
  x_0(0:psize-1) = xblock_{myid}
  do otherp = 1,m-1
    RECEIVE(x_0(psize*otherp),psize,otherp)
  end do
else
    SEND(xblock_{myid},psize,0)
endif

!   DISTRIBUTE_CENTRALIZED_TO_CYCLIC_PARTITIONED(x_0,xcyclic_{myid},m,psize)
if myid = 0 then
  xcyclic_{myid} = x_0(0:n-1:m)
  do otherp = 1,m-1
    SEND(x_0(otherp:n-1:m),psize,otherp)
  end do
else
    RECEIVE(xcyclic_{myid},psize,0)
endif

!   DISTRIBUTE_CYCLIC_PARTITIONED_TO_CENTRALIZED(xcyclic_{myid},x_0,m,psize)
if myid = 0 then
  x_0(0:n-1:m) = xcyclic_{myid}
  do otherp = 1,m-1
    RECEIVE(x_0(otherp:n-1:m),psize,otherp)
  end do
else
    SEND(xcyclic_{myid},psize,0)
endif
```

FIG. 5.8. *Code for Distribution Commands Occurring in Figure 5.7's Per–Processor Template Using Partitioned Data*

Figure 5.9, which is analogous to Figure 5.3. The direct implementation removes the centralized site as a bottleneck, but increases the number of messages from $2(m-1)$ to $m(m-1)$, as discussed in Section 5.1.3.

**6. Development of All–Processor Templates for Shared Memory Architectures with Local Memory Availability.**

```
    !  DISTRIBUTE_BLOCK_PARTITIONED_TO_CYCLIC_PARTITIONED(xblock_{myid},
                                              xcyclic_{myid},m,psize)
    xcyclic_{myid}(myid*psize/m:(myid+1)*psize/m-1) =
xblock_{myid}(myid:psize-1:m)
    do otherp = 0,m-1
      if otherp ≠ myid then
        SEND(xblock_{myid}(otherp:psize-1:m),psize/m,otherp)
        RECEIVE(xcyclic_{myid}(otherp*psize/m),psize/m,otherp)
      endif
    end do
```

Fig. 5.9. *Block–Partitioned to Cyclic–Partitioned Redistribution*

**6.1. Shared Memory Template.** The parallel computation plan of Figure 4.3 can be converted into a shared memory template by converting each of the two $p$ loops into a parallel loop, as shown in Figure 6.1. The template in Figure 6.1 is expressed in a style roughly based on OpenMP. We assume that there is a *thread* for each iteration of a parallel `do` loop, and that each thread can have data that is private to that thread. Moreover, we assume that each thread is assigned to a processor (with the possibility that multiple threads are assigned to any given processor), and data that is private to a given thread is stored in the local memory of the processor for that thread. In the template, we use the notational convention that `x` is a shared data array; and that scalar variables, such as `q`, are private to each thread. We also assume that each thread $p$ has a private array $\text{weight}_p$ for the weights used in the execution of that thread. Within each thread, the assignments to scalar variables and to elements of $\text{weight}_p$ modify private data. The assignment to elements of shared array `x` modify shared data, but because we are starting with the computation plan of Figure 4.3, the elements of `x` accessed by the various threads of a given `parallel do` loop are disjoint.

**6.2. Shared Memory Template Using Private Local Memory.** In Figure 6.1, all the accesses to `x` are to a shared data vector. The template can be modified so that the butterfly operates on private data, as shown in the all–processor template of Figure 6.2, which uses both a shared memory array `x`, and a private memory array $\text{x}_p$. Figure 6.3 shows the details of how the data copying is done. The OpenMP–style all–processor template uses assignment statements within each thread to copy data between shared memory and the private memory of that thread.

Figure 6.2 is an all–processor analogue of the per–processor distributed code template shown in Figure 5.6. The data copying via assignment statements in Figure 6.3 is analogous to the data copying via message–passing in Figures 5.2, 5.4, and 5.5. An OpenMP–style all–processor template permits a given thread to copy data between shared memory and private memory of that thread, but does not permit direct copying of data from the private memory of one thread to the private memory of a different thread. Accordingly, the DISTRIBUTE_BLOCK_TO_CYCLIC command from Figure 5.6 must be done as two commands (private block to centralized at the end of the first `parallel do`, followed by centralized to private block at the beginning of the second `parallel do`) in Figure 6.2, corresponding to the data movement in Figure 5.4, rather than that in Figure 5.3.

**6.3. Shared Memory Template Using Partitioned Data and Contiguous Access.** Figure 6.4 shows an OpenMP–style all–processor template for a shared

```
      parallel do p = 0,m-1
        do q = 1,breakpoint - 1
          L = 2**q
          do row = 0,L/2-1
            weight_p(row) = EXP((2*pi*i*row)/L)
          end do
          do col' = p*psize,(p+1)*psize-1,L
            do row = 0,L/2-1
              c = weight_p(row)*x(col'+row+L/2)
              d = x(col'+row)
              x(col'+row) = d + c
              x(col'+row+L/2) = d - c
            end do
          end do
        end do
      end parallel do
      parallel do p = 0,m-1
        do q = breakpoint,t
          L = 2**q
          do row = 0,L/2-1
            weight_p(row) = EXP((2*pi*i*row)/L)
          end do
          do col' = 0,n-1,L
            do row = p,L/2-1,m
              c = weight_p(row)*x(col'+row+L/2)
              d = x(col'+row)
              x(col'+row) = d + c
              x(col'+row+L/2) = d - c
            end do
          end do
        end do
      end parallel do
```

FIG. 6.1. *Simple Shared Memory All–Processor Template Obtained from Plan of Figure 4.3*

memory, based on the plan of Figure 4.9. We assume that there is a thread for each processor, and data that is private to a given thread is stored in the local memory of the processor for that thread. In the template, we use the notational convention that x is a shared data array; that $\text{weight}_p$, $\text{weightcyclic}_p$, $\text{xblock}_p$, and $\text{xcyclic}_p$ denote private data arrays for thread $p$; and that scalar variables are private to each thread. We assume that a given thread can access only shared data and its private data. Consequently, the block–to–cyclic redistribution cannot be done by moving data directly from a private array in one thread into a private array in a different thread. In Figure 6.4, the block–to–cyclic redistribution is done in two steps. In the first step, each thread of the first $p$ loop does a block–to–centralized copying of data from its private array $\text{xblock}_p$ into shared array x. Consequently, when the first $p$ loop concludes, the data from all the private $\text{xblock}_p$ arrays have been copied into shared array x. In the second step, each thread of the second $p$ loop does a centralized–to–block copying of data from x into its private array $\text{xcyclic}_p$.

```
  parallel do p = 0,m-1
    COPY_CENTRALIZED_TO_PRIVATE_BLOCK(x,x_p,p,psize)
    do q = 1,breakpoint - 1
      L = 2**q
      do row = 0,L/2-1
        weight_p(row) = EXP((2*pi*i*row)/L)
      end do
      do col' = p*psize,(p+1)*psize-1,L
        do row = 0,L/2-1
          c = weight_p(row)*x_p(col'+row+L/2)
          d = x_p(col'+row)
          x_p(col'+row) = d + c
          x_p(col'+row+L/2) = d - c
        end do
      end do
    end do
    COPY_PRIVATE_BLOCK_TO_CENTRALIZED(x_p,x,p,psize)
  end parallel do
  parallel do p = 0,m-1
    COPY_CENTRALIZED_TO_PRIVATE_CYCLIC(x,x_p,p,m,psize)
    do q = breakpoint,t
      L = 2**q
      do row = 0,L/2-1
        weight_p(row) = EXP((2*pi*i*row)/L)
      end do
      do col' = 0,n-1,L
        do row = p,L/2-1,m
          c = weight_p(row)*x_p(col'+row+L/2)
          d = x_p(col'+row)
          x_p(col'+row) = d + c
          x_p(col'+row+L/2) = d - c
        end do
      end do
    end do
    COPY_PRIVATE_CYCLIC_TO_CENTRALIZED(x_p,x,p,m,psize)
  end parallel do
```

FIG. 6.2. *All–Processor Shared and Private Memory Template Based on Template of Figure 6.1*

Figure 6.5 shows the details of the assignment statements that do the data copying between shared memory and the private memory of each thread.

**7. Preliminary Experimental Results.** There are various commercial and vendor specific libraries which include the FFT. The Numerical Analysis Group (NAG) and Visual Numerics (IMSL) provide and support finely tuned scientific libraries specific to various HPC platforms, e.g SGI and IBM. The SGI and IBM libraries, SCSL and ESSL, are even more highly tuned, due to their knowledge of proprietary information specific to their respective architectures. We ran experiments comparing our code to that in some of these libraries. We also did comparisons with the FFTW, although these results are harder to interpret because of the planning phase that is

```
    !   COPY_CENTRALIZED_TO_PRIVATE_BLOCK(x,x_p,p,psize)
    x_p(psize*p:psize*(p+1)-1:1) = x(psize*p:psize*(p+1)-1:1)

    !   COPY_PRIVATE_BLOCK_TO_CENTRALIZED(x_p,x,p,psize)
    x(psize*p:psize*(p+1)-1:1) = x_p(psize*p:psize*(p+1)-1:1)

    !   COPY_CENTRALIZED_TO_PRIVATE_CYCLIC(x,x_p,p,m,psize)
    x_p(p:n-1:m) = x(p:n-1:m)

    !   COPY_PRIVATE_CYCLIC_TO_CENTRALIZED(x_p,x,p,m,psize)
    x(p:n-1:m) = x_p(p:n-1:m)
```

FIG. 6.3. *Data Copying for All–Processor Shared and Private Memory Template of Figure 6.2*

part of the FFTW.

Our experiments are reported in [14, 15, 51]. Here, we give results from [15], where more details are available.

**7.1. Experimental Environment.** Our experiments were run on two systems:
1. A SGI/Cray Origin2000 at NCSA[8] in Illinois, with 48, 195Mhz R10000 processors, and 14GB of memory. The L1 cache size is 64KB (32KB Icache and 32 KB Dcache). The Origin2000 has a 4MB L2 cache. The OS is IRIX 6.5.
2. An IBM SP–2 at the MAUI High Performance Computing Center[9], with 32 P2SC 160Mhz processors, and 1 GB of memory. The L1 cache size is 160KB (32KB Icache and 128KB Dcache), and there is no L2 cache. The OS is AIX 4.3.

We tested the math libraries available on both machines: on the Origin 2000, IMSL Fortran Numerical Libraries version 3.01, NAG version Mark 19, and SGI's SCSL library; and on the SP–2, IBM's ESSL library. We also ran the FFTW on both machines.

**7.2. Experiments.** Experiments on the Origin 2000 were run using `bsub`, SGI's batch processing environment. Similarly, experiments on the SP–2 were run using the `loadleveler` batch processing environment. In both cases we used dedicated networks and processors. For each vector size ($2^3$ to $2^{24}$), experiments were repeated a minimum of three times and averaged. For improved optimizations, vendor compilers were used with the `-O3` and `-Inline` flags. We used Perl scripts to automatically compile, run, and time all experiments, and to plot our results for various problem sizes.

**7.3. Evaluation of Results.** Our results on the Origin 2000 and the SP–2, as shown in Figures 7.1 and 7.2, and Tables 7.1 and 7.2, indicate a performance improvement over standard libraries for many problem sizes, depending on the particular library.

**7.3.1. Origin 2000 Results.** Performance results for our monolithic FFT, which we call here *FFT-UA*, indicate a doubling of time when the vector size is doubled, for all vector sizes tried. IMSL doubled its performance up to $2^{19}$. At $2^{19}$

---
[8]This work was partially supported by National Computational Science Alliance, and utilized the NCSA SGI/CRAY Origin2000

[9]We would like to thank the Maui High Performance Computing Center for access to their IBM SP–2.

```
parallel do p = 0,m-1
  COPY_CENTRALIZED_TO_PRIVATE_BLOCK_PARTITIONED(x,xblock_p,m,psize,p)
  do q = 1,breakpoint - 1
    L = 2**q
    do row = 0,L/2-1
      weight_p(row) = EXP((2*pi*i*row)/L)
    end do
    do col'' = 0,psize-1,L
      do row = 0,L/2-1
        c = weight_p(row)*xblock_p(col''+row+L/2)
        d = xblock_p(col''+row)
        xblock_p(col''+row) = d + c
        xblock_p(col''+row+L/2) = d - c
      end do
    end do
  end do
  COPY_PRIVATE_BLOCK_PARTITIONED_TO_CENTRALIZED(xblock_p,x,p,m,psize,n)
end parallel do
parallel do p = 0,m-1
  COPY_CENTRALIZED_TO_PRIVATE_CYCLIC_PARTITIONED(x,xcyclic_p,m,psize,p,n)
  do q = breakpoint,t
    L = 2**q
    do row' = 0,L/(2*m)-1,1
      weightcyclic_p(row') = EXP((2*pi*i*(m*row'+p))/L)
    end do
    do col'' = 0,psize-1,L/m
      do row' = 0,L/(2*m)-1,1
        c = weightcyclic_p(row')*xcyclic_p(col''+row'+L/(2*m))
        d = xcyclic_p(col''+row')
        xcyclic_p(col''+row') = d + c
        xcyclic_p(col''+row'+L/(2*m)) = d - c
      end do
    end do
  end do
  COPY_PRIVATE_CYCLIC_PARTITIONED_TO_CENTRALIZED(xcyclic_p,x,m,psize,p,n)
end parallel do
```

FIG. 6.4. *All–Processor Shared Memory and Private Partitioned Data Template Obtained from Combined Plan of Figure 4.9*

there is a 400% degradation in performance, presumably due to a change in the use of the memory hierarchy. For NAG this degradation begins at $2^{18}$. The SGI library (SCSL) does not exhibit this degradation. SCSL may be doing machine specific optimizations, perhaps using more sophisticated out of core techniques similar to those described by Cormen [53], as evidenced by nearly identical performance times for $2^{17}$ and $2^{18}$.

**7.3.2. SP–2 Results.** *FFT-UA* outperforms ESSL for vector sizes up to $2^{14}$, except for two cases. For $2^{15}$ and $2^{16}$, the performance is slightly worse. ESSL does increasingly better as the problem size increases. The *FFT-UA* times continue to

```
!   COPY_CENTRALIZED_TO_PRIVATE_BLOCK_PARTITIONED(x,xblock_p,p,psize)
    xblock_p = x(psize*p:psize*(p+1)-1:1)

!   COPY_PRIVATE_BLOCK_PARTITIONED_TO_CENTRALIZED(xblock_p,m,psize,p)
    x(psize*p:psize*(p+1)-1:1) = xblock_p

!   COPY_CENTRALIZED_TO_PRIVATE_CYCLIC_PARTITIONED(x,xcyclic_p,m,psize,p,n)
    xcyclic_p = x(p:n-1:m)

!   COPY_PRIVATE_CYCLIC_PARTITIONED_TO_CENTRALIZED(xcyclic_p,x,m,psize,p,n)
    x(p:n-1:m) = xcyclic_p
```

Fig. 6.5. *Data Copying for All–Processor Shared Memory and Private Partitioned Data Template of Figure 6.4*

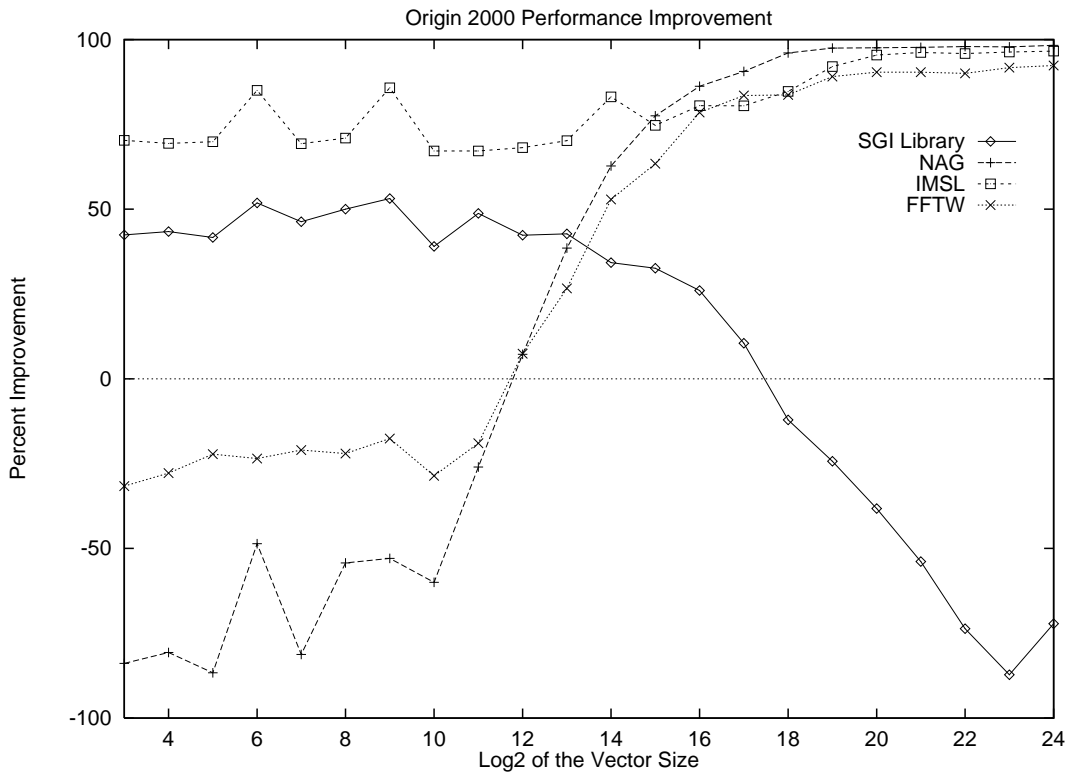

Fig. 7.1. *Percent improvement of* FFT-UA *over library code and FFTW on Origin 2000*

double from $2^{17}$ to $2^{23}$, reflecting the uniformity and simplicity of the code. The *FFT-UA* code is machine–independent, and relies on the efficiency of the compiler used. Presumably, the ESSL code is tuned to the machine architecture, using machine specific optimizations, and should be expected to perform better.

The ESSL code failed for problem sizes $2^{22}$ and higher, whereas *FFT-UA* successfully processed problem sizes through $2^{23}$, four times larger than the maximum handled by ESSL. The FFTW code failed for problem sizes $2^{23}$ and higher.

| Origin 2000 |||||
| Execution Time in Seconds |||||
| Size | FFT-UA | IMSL | NAG | SCSL |
|---|---|---|---|---|
| $2^3$ | 0.019 | 0.064 | 0.010 | 0.065 |
| $2^4$ | 0.018 | 0.061 | 0.010 | 0.047 |
| $2^5$ | 0.018 | 0.062 | 0.010 | 0.065 |
| $2^6$ | 0.017 | 0.116 | 0.011 | 0.073 |
| $2^7$ | 0.019 | 0.063 | 0.010 | 0.068 |
| $2^8$ | 0.018 | 0.062 | 0.011 | 0.105 |
| $2^9$ | 0.017 | 0.122 | 0.011 | 0.069 |
| $2^{10}$ | 0.021 | 0.065 | 0.013 | 0.056 |
| $2^{11}$ | 0.021 | 0.064 | 0.016 | 0.058 |
| $2^{12}$ | 0.021 | 0.067 | 0.023 | 0.067 |
| $2^{13}$ | 0.022 | 0.075 | 0.036 | 0.065 |
| $2^{14}$ | 0.024 | 0.144 | 0.065 | 0.066 |
| $2^{15}$ | 0.030 | 0.120 | 0.135 | 0.110 |
| $2^{16}$ | 0.040 | 0.209 | 0.296 | 0.080 |
| $2^{17}$ | 0.065 | 0.335 | 0.696 | 0.072 |
| $2^{18}$ | 0.126 | 0.829 | 3.205 | 0.075 |
| $2^{19}$ | 0.238 | 3.007 | 9.538 | 0.096 |
| $2^{20}$ | 0.442 | 9.673 | 18.40 | 0.143 |
| $2^{21}$ | 0.884 | 23.36 | 38.93 | 0.260 |
| $2^{22}$ | 1.910 | 46.70 | 92.75 | 0.396 |
| $2^{23}$ | 4.014 | 109.4 | 187.7 | 0.671 |
| $2^{24}$ | 7.550 | 221.1 | 442.7 | 1.396 |

TABLE 7.1
*Real Execution Times of* FFT-UA *and Comparative Libraries on Origin 2000*

**7.4. Conclusions from Experiments.** As evidence of the potential value of the uniform program design methodology outlined here, we constructed a machine–independent portable solution to the complex problem of faster and bigger FFTs. Reproducible experiments indicate that our designs outperform IMSL in all cases, NAG for sizes greater than $2^{11}$, SCSL for sizes less than $2^{18}$, and ESSL in some cases.

Our *single, portable, scalable* executable of approximately 2,600 bytes also must be compared with the large suite of machine–specific software required by NAG, IMSL, SCSL, and ESSL. The user of these machine–specific libraries must have a deep understanding of the package itself and the system on which it runs. Many of these packages require numerous parameters: fifteen for ESSL, and eight for SCSL. An incorrect decision for these parameters can result in poor performance and even possibly incorrect results. In comparison, a naive user of *FFT-UA* need only know the vector size on any platform.

**8. Conclusions, Extensions, and Future Research.** We have outlined a semi–mechanical methodology for developing efficient scientific software, centered on interactively developed sequences of algorithm transformations. We systematically improved the efficiency of the sequential expression of the high–level specification of the FFT algorithm, and formulated processor mapping and data decompositions strategies. As a phase of this methodology, abstract plans were constructed to specify which computations are to be done by each processor. Subsequently, templates were created attuned to various high–level architectural paradigms, e.g. shared and distributed memory multiprocessor computations. Parallel/distributed programs can be easily built from templates. Experimental comparisons indicate that our programs can be competitive in performance with software that has been hand–tuned to particular target architectures.

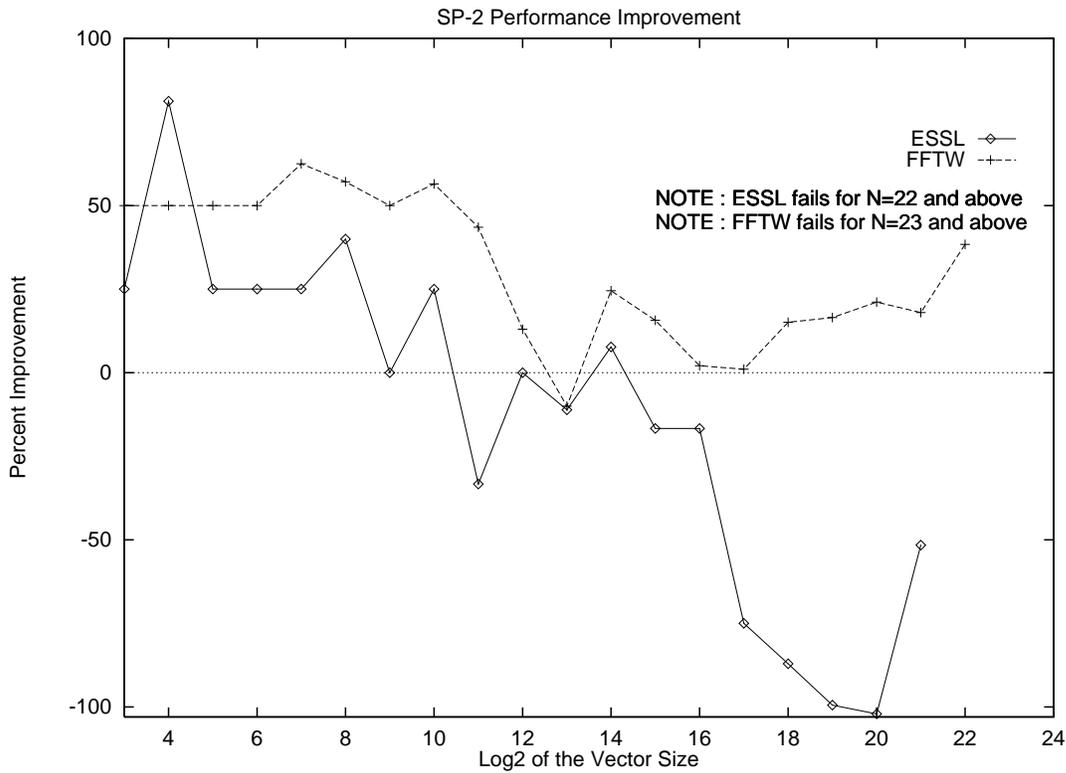

FIG. 7.2. *Percent improvement of* FFT-UA *over ESSL and FFTW on IBM SP–2*

| IBM SP–2 Execution Time in Seconds | | | | | |
|---|---|---|---|---|---|
| Size | FFT-UA | ESSL | Size | FFT-UA | ESSL |
| $2^3$ | 0.010 | 0.013 | $2^{14}$ | 0.040 | 0.043 |
| $2^4$ | 0.010 | 0.053 | $2^{15}$ | 0.070 | 0.060 |
| $2^5$ | 0.010 | 0.013 | $2^{16}$ | 0.140 | 0.120 |
| $2^6$ | 0.010 | 0.013 | $2^{17}$ | 0.280 | 0.160 |
| $2^7$ | 0.010 | 0.013 | $2^{18}$ | 0.580 | 0.310 |
| $2^8$ | 0.010 | 0.016 | $2^{19}$ | 1.216 | 0.610 |
| $2^9$ | 0.010 | 0.010 | $2^{20}$ | 2.580 | 1.276 |
| $2^{10}$ | 0.010 | 0.013 | $2^{21}$ | 5.553 | 3.663 |
| $2^{11}$ | 0.013 | 0.010 | $2^{22}$ | 12.12 | Failed |
| $2^{12}$ | 0.020 | 0.020 | $2^{23}$ | 25.25 | Failed |
| $2^{13}$ | 0.033 | 0.030 | | | |

TABLE 7.2
*Real Execution Times of* FFT-UA *and ESSL on IBM SP–2*

The algorithm variants developed here can be further optimized in several ways, consistent with our underlying methodology. Examples include tuning for shared memory, retiling to match the memory hierarchy, and better overlap of computation and communication. More generally, additional compiler–like analysis, transformations, and optimizations can be used to improve performance. This can be done during the various phases of the methodology. Finally, our proposed methodology for software development can be enhanced by a script–based mechanism for experimentally exploring the efficiency of alternative implementations for different ranges of problem sizes and different computing environments. The results of such experiments should prove useful in evaluating and guiding successive and/or alternative refinements.

The approach we have presented is similar in spirit to other efforts using libraries of algorithm building blocks based on C++ template classes such as POOMA. Based on the formalism of A Mathematics of Arrays (MoA) and an indexing calculus (i.e. the $\psi$ calculus) our approach enables the programmer to develop algorithms using high-level compiler-like optimizations through the ability to algebraically compose and reduce sequences of array operations. The resulting ONF is a prescription for code generation that can be directly translated into efficient code in the language of the programmer's choice be it for hardware or software application.

**Appendices:**

**Appendix A. Elements of the theory.**

**A.1. Indexing and Shapes.** The central operation of MoA is the indexing function

$$p\psi A$$

in which a vector of $n$ integers $p$ is used to select an item of the $n$-dimensional array $A$. The operation is generalized to select a partition of $A$, so that if $q$ has only $k < n$ components then

$$q\psi A$$

is an array of dimensionality $n - k$ and $q$ selects among the possible choices for the first $k$ axes. In MoA zero origin indexing is assumed. For example, if $A$ is the 3 by 5 by 4 array

$$\begin{bmatrix} 0 & 1 & 2 & 3 \\ 4 & 5 & 6 & 7 \\ 8 & 9 & 10 & 11 \\ 12 & 13 & 14 & 15 \\ 16 & 17 & 18 & 19 \end{bmatrix} \begin{bmatrix} 20 & 21 & 22 & 23 \\ 24 & 25 & 26 & 27 \\ 28 & 29 & 30 & 31 \\ 32 & 33 & 34 & 35 \\ 36 & 37 & 38 & 39 \end{bmatrix} \begin{bmatrix} 40 & 41 & 42 & 43 \\ 44 & 45 & 46 & 47 \\ 48 & 49 & 50 & 51 \\ 52 & 53 & 54 & 55 \\ 56 & 57 & 58 & 59 \end{bmatrix}$$

then

$$<1>\psi A = \begin{bmatrix} 20 & 21 & 22 & 23 \\ 24 & 25 & 26 & 27 \\ 28 & 29 & 30 & 31 \\ 32 & 33 & 34 & 35 \\ 36 & 37 & 38 & 39 \end{bmatrix}$$

$$<2\ 1>\psi A \ = \ <\ 44\ 45\ 46\ 47\ >$$

$$< 2\ 1\ 3 > \psi A\ =\ 47$$

Most of the common array manipulation operations found in languages like Fortran90, Matlab, ZPL, etc., can be defined from $\psi$ and a few elementary vector operations.

We now introduce notation to permit us to define $\psi$ formally and to develop the *Psi Correspondence Theorem* [54], which is central to the effective exploitation of MoA in array computations. We will use $A, B, ...$ to denote an array of numbers (integer, real, or complex). An array's dimensionality will be denoted by $d_A$ and will be assumed to be $n$ if not specified.

The shape of an array $A$, denoted by $s_A$, is a vector of integers of length $d_A$, each item giving the length of the corresponding axis. The total number of items in an array, denoted by $t_A$, is equal to the product of the items of the shape. The subscripts will be omitted in contexts where the meaning is obvious.

A full index is a vector of $n$ integers that describes one position in an $n$-dimensional array. Each item of a full index for $A$ is less than the corresponding item of $s_A$. There are precisely $t_A$ indices for an array(due to a zero index origin). A partial index of $A$ is a vector of $0 \leq k < n$ integers with each item less than the corresponding item of $s_A$.

We will use a tuple notation (omitting commas) to describe vectors of a fixed length. For example,

$$< i\ j\ k >$$

denotes a vector of length three. $<>$ will denote the empty vector which is also sometimes written as $\Theta$.

For every $n$-dimensional array $A$, there is a vector of the items of $A$, which we denote by the corresponding lower case letter, here $a$. The length of the vector of items is $t_A$. A vector is itself a one-dimensional array, whose shape is the one-item vector holding the length. Thus, for $a$, the vector of items of $A$, the shape of $a$ is

$$s_a\ =\ < t_A >$$

and the number of items or total number of components $a$[10] is

$$t_a\ =\ t_A.$$

The precise mapping of $A$ to $a$ is determined by a one-to-one ordering function, *gamma*. Although the choice of ordering is arbitrary, it is essential in the following that a specific one be assumed. By convention we assume the items of $A$ are placed in $a$ according to the lexicographic ordering of the indices of $A$. This is often referred to as *row major ordering*. Many programming languages lay out the items of multidimensional arrays in memory in a contiguous segment using this ordering. Fortran uses the ordering corresponding to a transposed array in which the axes are reversed *column major*. Scalars are introduced as arrays with an empty shape vector.

There are two equivalent ways of describing an array $A$:
**(1)** by its shape and the vector of items, i.e. $A = \{s_A, a\}$, or

---

[10]We also use $\tau a, \delta a$, and $\rho a$ to denote total number of components, dimensionality and shape of a.

**(2)** by its shape and a function that defines the value at every index $p$.

These two forms have been shown to be formally equivalent [55]. We wish to use the second form in defining functions on multidimensional arrays using their Cartesian coordinates (indices). The first form is used in describing address manipulations to achieve effective computation.

To complete our notational conventions, we assume that $p, q, ...$ will be used to denote indices or partial indices and that $u, v, ..$ will be used to denote arbitrary vectors of integers. In order to describe the $i_{th}$ item of a vector $a$, either $a_i$ or $a[i]$ will be used. If $u$ is a vector of $k$ integers all less than $t_A$, then $a[u]$ will denote the vector of length $k$, whose items are the items of $a$ at positions $u_j$, $j = 0, ..., k-1$.

Before presenting the formal definition of the $\psi$ indexing function we define a few functions on vectors:

| | |
|---|---|
| $u \mathbin{+\mkern-10mu+} v$ | catenation of vectors $u$ and $v$ |
| $u + v$ | itemwise vector addition assuming $t_u = t_v$ |
| $u \times v$ | itemwise vector multiplication |
| $n + u, u + n$ | addition of a scalar to each item of a vector |
| $n \times u, u \times n$ | multiplication of each item of a vector by a scalar |
| $\iota\, n$ | the vector of the first n integers starting from 0 |
| $\pi\, v$ | a scalar which is the product of the components of v |
| $k\, \triangle\, u$ | when $k \geq 0$ the vector of the first $k$ items of $u$,(called *take*) and when $k < 0$ the vector of the last $k$ items of $u$ |
| $k\, \triangledown\, u$ | when $k \geq 0$ the vector of $t_u - k$ last items of $u$,(called *drop*) and when $k < 0$ the vector of the first $t_u - |k|$ items of $u$ |
| $k\, \theta\, u$ | when $k \geq 0$ the vector of $(k\, \triangledown\, u) \mathbin{+\mkern-10mu+} (k\, \triangle\, u)$ and when $k < 0$ the vector or $(k\, \triangle\, u) \mathbin{+\mkern-10mu+} (k\, \triangledown\, u)$ |

DEFINITION A.1. *Let A be an n-dimensional array and p a vector of integers. If p is an index of A,*

$$p \psi A = a[\gamma(s_A, p)],$$

*where*

$$\gamma(s_A, p) = x_{n-1} \quad \text{defined by the recurrence}$$
$$x_0 = p_0,$$
$$x_j = x_{j-1} * s_j + p_j, \quad j = 1, ..., n-1.$$

*If p is partial index of length $k < n$,*

$$p \psi A = B$$

*where the shape of B is*

$$s_B = k\, \triangledown\, s_A,$$

*and for every index q of B,*

$$q \psi B = (p \mathbin{+\mkern-10mu+} q) \psi A$$

The definition uses the second form of specifying an array to define the result of a partial index. For the index case, the function $\gamma(s, p)$ is used to convert an index $p$ to an integer giving the location of the corresponding item in the row major order list of items of an array of shape $s$. The recurrence computation for $\gamma$ is the one used in most compilers for converting an index to a memory address [56].

COROLLARY A.2. $<>\psi\ A = A$.

The following theorem shows that a $\psi$ selection with a partial index can be expressed as a composition of $\psi$ selections.

THEOREM A.3. *Let A be an n-dimensional array and p a partial index so that $p = q +\!\!+ r$. Then*

$$p\psi A = r\psi(q\psi A).$$

*Proof:* The proof is a consequence of the fact that for vectors $u,v,w$

$$(u +\!\!+ v) +\!\!+ w = u +\!\!+ (v +\!\!+ w).$$

If we extend $p$ to a full index by $p +\!\!+ p'$, then

$$\begin{aligned}
p'\psi(p\psi A) &= (p +\!\!+ p')\psi A \\
&= ((q +\!\!+ r) +\!\!+ p')\psi A \\
&= (q +\!\!+ (r +\!\!+ p'))\psi A \\
&= (r +\!\!+ p')\psi(q\psi A) \\
&= p'\psi(r\psi(q\psi A)) \\
p\psi A &= r\psi(q\psi A)
\end{aligned}$$

which completes the proof.

We can now use *psi* to define other operations on arrays. For example, consider definitions of *take* and *drop* for multidimensional arrays.

DEFINITION A.4 (take: $\triangle$). *Let A be an n-dimensional array, and k a non-negative integer such that $0 \leq k < s_0$. Then*

$$k\ \triangle\ A = B$$

where

$$s_B = <k> +\!\!+ (1\ \triangledown\ s_A)$$

and for every index $p$ of $B$,

$$p\psi B = p\psi A.$$

(In MoA $\triangle$ is also defined for negative integers and is generalized to any vector u with its absolute value vector a partial index of A. The details are omitted here.)

DEFINITION A.5 (reverse: $\Phi$). *Let A be an n-dimensional array. Then*

$$s_{\Phi A} = s_A$$

and for every integer $i$, $0 \leq i < s_0$,

$$<i>\psi\Phi A = <s_0 - i - 1>\psi A.$$

This definition of $\Phi$ does a reversal of the 0th axis of A.

Note also that all operations are over the 0th axis. The operator $\Omega$ [26] extends operations over all other dimensions.

**Example.** Consider the evaluation of the following expression using the 3 by 5 by 4 array, $A$, introduced in Section A.1.

$$< 1\ 2 > \psi(2\ \triangle\ \Phi A) \tag{A.1}$$

where A is the array given in the previous section. The shape of the result is

$$\begin{aligned}
&2\ \triangledown\ s_{(2\triangle \Phi A)} \\
&= 2\ \triangledown\ (<2> \mathbin{+\!\!+} (1\ \triangledown\ s_{\Phi A})) \\
&= 2\ \triangledown\ (<2> \mathbin{+\!\!+} (1\ \triangledown\ s_A)) \\
&= 2\ \triangledown\ (<2> \mathbin{+\!\!+} <5\ 4>) \\
&= 2\ \triangledown\ <2\ 5\ 4> \\
&= <4>\,.
\end{aligned} \tag{A.2}$$

The expression can be simplified using the definitions:

$$\begin{aligned}
&<1\ 2> \psi(2\ \triangle\ \Phi A) \\
&= <1\ 2> \psi \Phi A \\
&= <2> \psi(<1> \psi \Phi A) \\
&= <2> \psi(<3-1-1> \psi A) \\
&= <1\ 2> \psi A
\end{aligned} \tag{A.3}$$

This process of simplifying the expression for the item in terms of its Cartesian coordinates is called *Psi Reduction*. The operations of MoA have been designed so that all expressions can be reduced to a minimal normal form [26]. Some MoA operations defined by *psi* are found in Fig.A.1.

**Appendix B. Higher Order Operations.**

Thus far operation on arrays, such as concatenation, rotation, etc., have been performed over their 0th dimensions. We introduce the higher order binary operation $\Omega$, which is defined when its left argument is a unary or binary operation and its right argument is a vector describing the dimension upon which operations are to be performed, or which sub-arrays are used in operations. The dimension upon which operations are to be performed is often called the *axis* of operation. The result of $\Omega$ is a unary or binary operation.

**B.1. Definition of $\Omega$.** $\Omega$ is defined whenever its left argument is a unary or binary operation, $f$ or $g$ respectively ($f$ and $g$ include the outcome of higher order operation). $\Omega$'s right argument is a vector, $\vec{d}$, such that $\rho \vec{d} \equiv <1>$ or $\rho \vec{d} \equiv <2>$ depending on whether the operation is unary or binary. Commonly, $f$ (or $g$) will be an operation which determines the shape of its result based on the shapes of its arguments, not on the values of their entries, i.e. for all appropriate arguments $\rho f \xi$ is determined by $\rho \xi$ and $\rho \xi_l g \xi_r$ is determined by $\rho \xi_l$ and $\rho \xi_r$.

*Definition 0:* ${}_f \Omega_{\vec{d}}$ is defined when $f$ is a one argument function, $\vec{d} \equiv <\sigma>$, with $\sigma \geq 0$.

For any non-empty array $\xi$,

$${}_f \Omega_{\vec{d}} \xi \tag{B.1}$$

| Symbol | Name | Description |
| --- | --- | --- |
| $\delta$ | Dimensionality | Returns the number of dimensions of an array. |
| $\rho$ | Shape | Returns a vector of the upper bounds or sizes of each dimension in an array. |
| $\iota \xi^n$ | Iota | When $n = 0$(scalar), returns a vector containing elements 0, to $\xi^0 - 1$. When $n = 1$(vector), returns an array of indices defined by the shape vector $\xi^1$ |
| $\psi$ | Psi | The main indexing function of the Psi Calculus which defines all operations in MoA. Returns a scalar if a full index is provided, a sub-array otherwise. |
| rav | Ravel | vectorizes a multi-dimensional array based on an array's layout($\gamma_{row}, \gamma_{col}, \gamma_{sparse}, ...$) |
| $\gamma$ | Gamma | Translates indices into offsets given a shape. |
| $\gamma'$ | Gamma Inverse | Translates offsets into indices given a shape. |
| $\vec{s} \,\widehat{\rho}\, \xi$ | Reshape | Changes the shape vector of an array, possibly affecting its dimensionality. Reshape depends on layout($\gamma$). |
| $\pi \vec{x}$ | Pi | Returns a scalar and is equivalent to $\prod_{i=0}^{(\tau x)-1} x[i]$ |
| $\tau$ | Tau | Returns the number of components in an array,($\tau\xi \equiv \pi(\rho\xi)$) |
| $\xi_l \,\text{+}\!\!\text{+}\, \xi_r$ | Catenate | Concatenates two arrays over their primary axis. |
| $\xi_l f \xi_r$ | Point-wise Extension | A data parallel application of $f$ is performed between all elements of the arrays. |
| $\sigma f \xi_r$ $\xi_l f \sigma$ | Scalar Extension | $\sigma$ is used with every component of $\xi_r$ in the data parallel application of $f$. |
| $\triangle$ | Take | Returns a sub-array from the beginning or end of an array based on its argument being positive or negative. |
| $\triangledown$ | Drop | The inverse of Take |
| $_{op}\text{red}$ | Reduce | Reduce an array's dimension by one by applying op over the primary axis of an array. |
| $\Phi$ | Reverse | Reverses the components of an array. |
| $\Theta$ | Rotate | Rotates, or shifts cyclically, components of an array. |
| $\bigcirc\!\!\!\!\backslash$ | Transpose | Transposes the elements of an array based on a given permutation vector |
| $\Omega$ | Omega | Applies a unary or binary function to array argument(s) given partitioning information. $\Omega$ is used to perform all operations (defined over the primary axis only) over all dimensions. |

Fig. A.1. *Summary of MoA Operations*

is defined provided (*i*) $(\delta\xi) \geq \sigma$, and provided certain other conditions, stated below, are met. Let

$$\vec{u} \equiv (-\sigma) \,\triangledown\, \rho\xi. \tag{B.2}$$

We can write

$$\rho\xi \equiv \vec{u} \,\text{+}\!\!\text{+}\, \vec{z} \tag{B.3}$$

where $\vec{z} \equiv (-\sigma) \,\triangle\, \rho\xi$.

We further require (*ii*) there exists $\vec{w}$ such that for $0 \leq^\star \vec{i} <^\star \vec{u}$,

$$f(\vec{i}\psi\xi) \tag{B.4}$$

is defined and has shape $\vec{w}$. The notation $0 \leq^\star \vec{i} <^\star \vec{u}$, is a shorthand which implies that we are comparing two vectors $\vec{i}$ and $\vec{u}$ component by component. With this

$$\rho(_f\Omega_{\vec{d}})\xi \equiv \vec{u} \,\text{+}\!\!\text{+}\, \vec{w} \tag{B.5}$$

and for $0 \leq^\star \vec{i} <^\star \vec{u}$,

$$\vec{i}\psi(_f\Omega_{\vec{d}})\xi \equiv f(\vec{i}\psi\xi) \tag{B.6}$$

*End of Definition 0*

Note that condition (*ii*) is easily satisfied for common $f$'s.

*Definition 1:* We similarly define $\Omega$ when its function argument is a binary operation $g$. $_g\Omega_{\vec{d}}$ is defined when $g$ is a two argument function, $\vec{d} \equiv <\sigma_l\ \sigma_r>$, with $\sigma_l \geq 0$, and $\sigma_r \geq 0$.

For any non-empty arrays, $\xi_l$, and $\xi_r$,

$$\xi_l(_g\Omega_{\vec{d}})\xi_r \tag{B.7}$$

is defined provided (*i*) $(\delta\xi_l) \geq \sigma_l$ and $(\delta\xi_r) \geq \sigma_r$, and provided certain other conditions, stated below, are met.

We let $\lfloor$ denote the binary operation minimum and let

$$m \equiv ((\delta\xi_l) - \sigma_l)\lfloor((\delta\xi_r) - \sigma_r). \tag{B.8}$$

We require that (*ii*) $((-m) \triangle (-\sigma_l) \triangledown \rho\xi_l) \equiv ((-m) \triangle (-\sigma_r) \triangledown \rho\xi_r)$.

Let

$$\vec{x} \equiv ((-m) \triangle (-\sigma_l) \triangledown \rho\xi_l) \equiv ((-m) \triangle (-\sigma_r) \triangledown \rho\xi_r), \tag{B.9}$$

$$\vec{u} \equiv (-m) \triangledown (-\sigma_l) \triangledown \rho\xi_l, \tag{B.10}$$

$$\vec{v} \equiv (-m) \triangledown (-\sigma_r) \triangledown \rho\xi_r. \tag{B.11}$$

Note that $\vec{u} \equiv <>$ or $\vec{v} \equiv <>$ (both could be empty). We can now write

$$\rho\xi_l \equiv \vec{u} \mathbin{+\!\!+} \vec{x} \mathbin{+\!\!+} \vec{y}, \tag{B.12}$$

and,

$$\rho\xi_r \equiv \vec{v} \mathbin{+\!\!+} \vec{x} \mathbin{+\!\!+} \vec{z} \tag{B.13}$$

where $\vec{y} \equiv (-\sigma_l) \triangle \rho\xi_l$ and $\vec{z} \equiv (-\sigma_r) \triangle \rho\xi_r$. Any of the vectors above could be empty.

We also require (*iii*) there exists a fixed vector $\vec{w}$ such that for $0 \leq^\star \vec{i} <^\star \vec{u}$, $0 \leq^\star \vec{j} <^\star \vec{v}$, $0 \leq^\star \vec{k} <^\star \vec{x}$,

$$((\vec{i} \mathbin{+\!\!+} \vec{k})\psi\xi_l)g((\vec{j} \mathbin{+\!\!+} \vec{k})\psi\xi_r) \tag{B.14}$$

is defined and has shape $\vec{w}$.

With all this

$$\rho(\xi_l(_g\Omega_{\vec{d}})\xi_r) \equiv \vec{u} \mathbin{+\!\!+} \vec{v} \mathbin{+\!\!+} \vec{x} \mathbin{+\!\!+} \vec{w} \tag{B.15}$$

and for $0 \leq^\star \vec{i} <^\star \vec{u}$, $0 \leq^\star \vec{j} <^\star \vec{v}$, $0 \leq^\star \vec{k} <^\star \vec{x}$,

$$(\vec{i} \mathbin{+\!\!+} \vec{j} \mathbin{+\!\!+} \vec{k})\psi\xi_l(_g\Omega_{\vec{d}})\xi_r \equiv ((\vec{i} \mathbin{+\!\!+} \vec{k})\psi\xi_l)g((\vec{j} \mathbin{+\!\!+} \vec{k})\psi\xi_r) \tag{B.16}$$

*End of Definition 1*

<div style="text-align:center">REFERENCES</div>

[1] W. Humphrey, S. Karmesin, F. Bassetti, and J. Reynders. Optimization of data–parallel field expressions in the POOMA framework. In Y. Ishikawa, R. R. Oldehoeft, J. Reynders, and M. Tholburn, editors, *Proc. First International Conference on Scientific Computing in Object–Oriented Parallel Environments (ISCOPE '97)*, volume 1343 of *Lecture Notes in Computer Science*, pages 185–194, Marina del Rey, CA, December 1997. Springer–Verlag.

[2] S. Karmesin, J. Crotinger, J. Cummings, S. Haney, W. Humphrey, J. Reynders, S. Smith, and T. Williams. Array design and expression evaluation in POOMA II. In D. Caromel, R. R. Oldehoeft, and M. Tholburn, editors, *Proc. Second International Symposium on Scientific Computing in Object–Oriented Parallel Environments (ISCOPE '98)*, volume 1505 of *Lecture Notes in Computer Science*, Santa Fe, NM, December 1998. Springer–Verlag.

[3] Marnix Vlot. The POOMA architecture. *j-LECT-NOTES-COMP-SCI*, 503:365–, 1991.

[4] Rogier Wester and Ben Hulshof. The POOMA operating system. *j-LECT-NOTES-COMP-SCI9*, 503:396–??, 1991.

[5] John V. W. Reynders, Paul J. Hinker, Julian C. Cummings, Susan R. Atlas, Subhankar Banerjee, William F. Humphrey, Steve R. Karmesin, Katarzyna Keahey, M. Srikant, and Mary-Dell Tholburn. POOMA. In Gregory V. Wilson and Paul Lu, editors, *Parallel Programming Using C++*. MIT Press, 1996.

[6] David R. Musser and Atul Saini. *STL Tutorial and Reference Guide*. Addison-Wesley, 1996.

[7] Matthew H. Austern. *Generic Programming and the STL: Using and Extending the C++ Standard Template Library*. Addison-Wesley, 1998.

[8] Mark Weiss. *Algorithms, Data Structures, and Problem Solving C++*. Addison-Wesley, 1996.

[9] John V. W. Reynders, Paul J. Hinker, Julian C. Cummings, Susan R. Atlas, Subhankar Banerjee, William F. Humphrey, Steve R. Karmesin, Katarzyna Keahey, M. Srikant, and Mary-Dell Tholburn. "POOMA". In *Parallel Programming Using C++*. The MIT Press, Cambridge, 1996.

[10] Todd Veldhuizen. *"Expression Templates." C++ Report 7:5 (June, 1995)*. Sigs Books, NY, 1995.

[11] A. Lumsdaine and B. McCandless. Parallel extensions to the matrix template library. In SIAM, editor, *Proceedings of the 8th SIAM Conference on Parallel Processing for Scientific Computing*, 1997.

[12] A. Lumsdaine. The matrix template library: A generic programming approach to high performance numerical linear algebra. In *Proceedings of International Symposium on Computing in Object-Oriented Parallel Environments*, 1998.

[13] Papers I and II of this series have been submitted to the Journal of Computational Physics but are available as online as
http://trr.albany.edu/documents/TR00004, and
http://trr.albany.edu/documents/TR00005, respectively.

[14] L. R. Mullin and S. G. Small. Three easy steps to a faster FFT (no, we don't need a plan). In M. S. Obaidat and F. Davoli, editors, *Proceedings of 2001 International Symposium on Performance Evaluation of Computer and Telecommunication Systems, SPECTS 2001*, pages 604–611, Orlando, FL, July 2001.

[15] L. R. Mullin and S. G. Small. Three easy steps to a faster FFT (the story continues . . .). In H. R. Arabnis, editor, *Proceedings of the International Conference on Imaging Science, Systems, and Technology, CISST 2001*, pages 107–113, Las Vegas, NV, June 2001.

[16] L. R. Mullin and S. G. Small. Four easy ways to a faster fft. *Journal of Mathematical Modelling and Algorithms*, 1:193, 2002.

[17] L. Mullin. A uniform way of reasoning about array–based computation in radar: Algebraically connecting the hardware/software boundary. *Digital Signal Processing (to appear)*, to appear.

[18] J. E. Raynolds and L. R. Mullin. *Computer Physics Communications*, 170:1, 2005.

[19] T. Cormen. Everything you always wanted to know about out–of–core FFTs but were aftaid to ask. COMPASS Colloquia Series, U Albany, SUNY, 2000.

[20] J.-W. Hong and H. T. Kung. I/o complexity: the red-blue pebble game., 1981.

[21] J. E. Savage. Entending the Hong-Kung model to memory hierarchies. *Lecture notes in Computer Science*, 959:270–281, 1995.

[22] Jeffrey Scott Vitter and Elizabeth A. M. Shriver. Algorithms for parallel memory II: Hierarchical multilevel memories. Technical Report Technical report DUKE–TR–1993-02, 1993.

[23] Lenore R. Mullin, Harry B Hunt III, and Daniel J. Rosenkrantz. A transformation-based approach for the design of parallel/distributed scientific software: the FFT. http://trr.albany.edu/documents/TR00002.

[24] W. Gropp, E. Lusk, and A. Skjellum. *Using MPI: Portable Parallel Programming with the*

[25] R. Chandra, R. Menon, L. Dagum, D. Kohr, D. Maydan, and J. McDonald. *Parallel Programming in OpenMP*. Morgan Kaufmann Publishers, Los Altos, CA, 2000.

[26] L. M. R. Mullin. *A Mathematics of Arrays*. PhD thesis, Syracuse University, December 1988.

[27] L. Mullin. The Psi compiler project. In *Workshop on Compilers for Parallel Computers*. TU Delft, Holland, 1993.

[28] L. R. Mullin, D. Dooling, E. Sandberg, and S. Thibault. Formal methods for scheduling, routing and communication protocol. In *Proceedings of the Second International Symposium on High Performance Distributed Computing (HPDC-2)*. IEEE Computer Society, July 1993.

[29] L. R. Mullin, D. Eggleston, L. J. Woodrum, and W. Rennie. The PGI–PSI project: Preprocessing optimizations for existing and new F90 intrinsics in HPF using compositional symmetric indexing of the Psi calculus. In M. Gerndt, editor, *Proceedings of the 6th Workshop on Compilers for Parallel Computers*, pages 345–355, Aachen, Germany, December 1996. Forschungszentrum Jülich GmbH.

[30] Douglas F. Elliott and K. Ramamohan Rao. *Fast Transforms: Algorithms, Analyses, Applications*. Academic Press, Inc., 1982.

[31] R. Tolimieri, M. An, and C. Lu. *Algorithms for Discrete Fourier Tranform and Convolution*. Springer-Verlag, 1989.

[32] R. Tolimieri, M. An, and C. Lu. *Mathematics of Multidimensional Fourier Transform Algorithms*. Springer-Verlag, 1993.

[33] D. L. Dai, S. K. S. Gupta, S. D. Kaushik, and J. H. Lu. EXTENT: A portable programming environment for designing and implementing high–performance block–recursive algorithms. In *Proceedings, Supercomputing '94*, pages 49–58, Washington, DC, November 1994. IEEE Computer Society Press.

[34] J. Granata, M. Conner, and R. Tolimieri. Recursive fast algorithms and the role of the tensor product. *IEEE Transactions on Signal Processing*, 40(12):2921–2930, December 1992.

[35] S. Gupta, C.-H. Huang, P. Sadayappan, and R. Johnson. On the synthesis of parallel programs from tensor product formulas for block recursive algorithms. In Uptal Banerjee, David Gelernter, Alex Nicolau, and David Padua, editors, *Proceedings of the 5th International Workshop on Languages and Compilers for Parallel Computing*, volume 757 of *Lecture Notes in Computer Science*, pages 264–280, New Haven, CT, August 1992. Springer-Verlag.

[36] S. K. S. Gupta, C.-H. Huang, P. Sadayappan, and R. W. Johnson. A framework for generating distributed–memory parallel programs for block recursive algorithms. *Journal of Parallel and Distributed Computing*, 34(2):137–153, May 1996.

[37] Jeremy Johnson, Robert W. Johnson, David A. Padua, and Jianxin Xiong. Searching for the best FFT formulas with the SPL compiler. In S. P. Midkiff et al., editors, *Proceedings of the 13th International Workshop on Languages and Compilers for Parallel Computing 2000 (LCPC 2000)*, volume 2017 of *Lecture Notes in Computer Science*, pages 112–126, Yorktown Heights, NY, August 2000. Springer.

[38] Jianxin Xiong, Jeremy Johnson, Robert Johnson, and David Padua. SPL: A language and compiler for DSP algorithms. In *Proceedings of the ACM SIGPLAN'01 Conference on Programming Language Design and Implementation*, pages 298–308, Snowbird, UT, 2001. ACM Press.

[39] M. Frigo and S. G. Johnson. The fastest Fourier transform in the West. Technical Report MIT-LCS-TR-728, Massachusetts Institute of Technology, September 1997.

[40] M. Frigo and S. G. Johnson. FFTW: An adaptive software architecture for the FFT. In *Proc. IEEE International Conf. on Acoustics, Speech, and Signal Processing, Vol. 3*, pages 1381–1384, May 1998.

[41] M. Frigo. A fast Fourier transform compiler. In *Proceedings of the ACM SIGPLAN '99 Conference on Programming Language Design and Implementation*, pages 169–180, Atlanta, GA, May 1999.

[42] Kang Su Gatlin and Larry Carter. Faster FFTs via architecture–cognizance. In *Proceedings of the 2000 International Conference on Parallel Architectures and Compilation Techniques (PACT '00)*, pages 249–260, Philadelphia, PA, October 2000. IEEE Computer Society Press.

[43] Dragan Mirković, Rishad Mahasoom, and Lennart Johnsson. An adaptive software library for fast Fourier transforms. In *Conference Proceedings of the 2000 International Conference on Supercomputing*, pages 215–224, Santa Fe, New Mexico, May 2000. ACM SIGARCH.

[44] R. C. Agarwal, F. G. Gustavson, and M. Zubair. A high performance parallel algorithm for 1-D FFT. In *Proceedings, Supercomputing '94*, pages 34–40, Washington, DC, November 1994. IEEE Computer Society Press.

[45] D. Culler, R. Karp, D. Patterson, A. Sahay, K. E. Schauser, E. Santos, R. Subramonian, and

T. von Eicken. LogP: Toward a realistic model of parallel computation. In *Proceedings of the Fourth ACM SIGPLAN Symposium on Principles and Practice of Parallel Programming*, pages 1–12, May 1993.

[46] Anshul Gupta and Vipin Kumar. The scalability of FFT on parallel computers. *IEEE Transactions on Parallel and Distributed Systems*, 4(8):922–932, August 1993.

[47] S. K. S. Gupta, C.-H. Huang, P. Sadayappan, and R. W. Johnson. Implementing fast Fourier transforms on distributed–memory multiprocessors using data redistributions. *Parallel Processing Letters*, 4(4):477–488, December 1994.

[48] Douglas Miles. Compute intensity and the FFT. In *Proceedings, Supercomputing '93*, pages 676–684, Portland, OR, November 1993. IEEE Computer Society Press.

[49] Pamela Thulasiraman, Kevin B. Theobald, Ashfaq A. Khokhar, and Guang R. Gao. Multi-threaded algorithms for fast Fourier transform. In *Proceedings of the 12th Annual ACM Symposium on Parallel Algorithms and Architecture (SPAA-00)*, pages 176–185, NY, July 2000. ACM Press.

[50] H. B. Hunt III, L. R. Mullin, and D. J. Rosenkrantz. Experimental construction of a fine–grained polyalgorithm for the FFT. In *Proceedings of the International Conference on Parallel and Distributed Processing Techniques and Applications (PDPTA'99)*, pages 1641–1647, Las Vegas, NV, June 1999.

[51] L. R. Mullin and S. G. Small. Four easy ways to a faster FFT. *Journal of Mathematical Modelling and Algorithms*, 1(3):193–214, 2002.

[52] Charles Van Loan. *Computational Frameworks for the Fast Fourier Transform*. Frontiers in Applied Mathematics. SIAM, 1992.

[53] Thomas H. Cormen and David M. Nicol. Performing out–of–core FFTs on parallel disk systems. *Parallel Computing*, 24(1):5–20, 1998.

[54] L. Mullin and M. Jenkins. Effective data parallel computation using the Psi calculus. *Concurrency – Practice and Experience*, September 1996.

[55] M. Jenkins, 94. Research Communications.

[56] P. Lewis, D. Rosenkrantz, and R. Stearns. *Compiler Design Theory*. Addison-Wesley, 1976.